\documentclass[a4,11pt]{article}
\usepackage{graphics}
\usepackage{amsmath}
\usepackage{amsfonts}
\usepackage{amssymb}
\usepackage{mathrsfs}
\usepackage{floatflt}
\newtheorem{lemma}{Lemma}
\newtheorem{corollario}{Corollary}
\newtheorem{theorem}{Theorem}
\newtheorem{definizione}{Definition}
\newtheorem{condizione}{Condition}

\def\min{{\rm min}}
 \def\eps{ { \varepsilon } }

\def\phi{{\varphi}}
\newcommand{\equal}{\buildrel {\rm def} \over {=} }
\newcommand{\vett}[1]{\mathbf{#1}} 
\newcommand{\T}{T}
\newcommand{\xgot}{{\mathfrak{x}}}
\newcommand{\indmono}{i}
\newcommand{\ind}{\vett \indmono}
\newcommand{\indamono}{j}
\newcommand{\inda}{\vett \indamono}
\newcommand{\indb}{\vett k}
\newcommand{\range}{r}

\newcommand{\dif}{\mathrm{d}}

\newcommand{\nucleo}{\Theta} 
\newcommand{\blank}{\vskip 1ex \noindent} 

\begin{document}

\title{Exponentially long  stability times for a nonlinear lattice in
       the thermodynamic limit}

\author{Andrea~Carati\thanks{Universit\`a di Milano, Dipartimento di
       Matematica, Via Saldini 50, 20133 
	 Milano, Italy. E-mail: \texttt{andrea.carati@unimi.it}} 
        \and
        Alberto~Mario~Maiocchi\footnotemark[1]
}

\date{\today}

\maketitle

\begin{abstract}
In this paper, we construct an adiabatic invariant for a large 1--$d$
lattice of particles, which is the so called Klein Gordon lattice. The time
evolution of such a quantity is bounded by a stretched exponential as the
perturbation parameters tend to zero. At variance  with  the results
available in the literature, our result holds uniformly in the
thermodynamic limit. The proof consists of two steps: first, one uses 
 techniques of Hamiltonian perturbation theory to construct a formal
 adiabatic invariant; second, one uses probabilistic
methods to show that, with large probability, the adiabatic
invariant is approximately constant. As a corollary, we can give a bound
from below to the relaxation time for the considered system, through
estimates on the autocorrelation of the adiabatic invariant.
\end{abstract}

\section{Introduction} 

One of the open problems of Hamiltonian perturbation theory is how to
extend to infinite dimensional systems, at a finite specific energy
(or temperature), the results known for systems 
with a finite number of degrees of freedom. Indeed there exist results
both for infinite systems such as partial 
differential equation (see, for example, \cite{pde1,pde2}) or
on infinite lattice systems (see  \cite{froehlich,enfinita}),
but only for a finite total energy of the sistem, i.e., at zero temperature.

In the present paper we provide perturbation estimates on
the so called Klein Gordon lattice in the thermodynamic limit, at a
finite temperature, by controlling, in place of the usual $L^\infty$
norm, the $L^2$ norm relative to the Gibbs measure. If we
denote by $H$ the Hamiltonian of the system and by
$\mathcal{M}$ the corresponding phase space, the Gibbs measure is defined by
\begin{equation}\label{eq:gibbs}
\mu(\dif x)\equal \frac{\exp(-\beta H(x))}{\mathcal{Z}(\beta)}\dif x\ ,
\end{equation}
where $ \mathcal{Z}(\beta) \equal  \int_\mathcal{M} \exp(-\beta H(x))\dif x $
is the partition function, and $\beta>0$ the inverse temperature.

We construct an adiabatic invariant whose time derivative
has an $L^2$ norm exponentially small in the perturbation parameters
(see Theorem~\ref{teor:main}). The construction of the adiabatic
invariant is standard (see~\cite{giorgilli}), but the estimate of its
time derivative in the $L^2$ norm involves some probabilistic
techniques, which have been developed in the frame of statistical
mechanics. In fact, since $L^2$ is not a Banach algebra, the usual
scheme of perturbation estimates cannot be implemented. So the use of
the algebra property is replaced here by a control of the decay of
spatial correlations between the sites of the lattice, making use of
techniques introduced by Dobrushin (see \cite{do1}). In particular, we are
able to show that, for lattices in any dimension with finite range
interaction (i.e., in which each
particle interacts only with a finite numbers of neighbouring ones) the spatial
correlations decay exponentially fast with the distance. This requires also
an estimate on the marginal probability densities induced by the 
measure $\mu$ on subsystems of finite size: this is done by adapting
to lattices the techniques introduced by Bogolyubov et al. (see
\cite{bogoljubov}) in interacting gas theory.

The paper is organized as follows. The main result
on the considered model, namely the construction of an adiabatic
invariant in the thermodynamic limit, is stated in Section~\ref{sez:stab}
(Theorem~\ref{teor:main}), togheter with two corollaries concerning
a control on the time evolution of the adiabatic invariant and a lower
bound to its time autocorrelation.  Then, in Section~\ref{sez:schema},
we present the scheme of the proof of Theorem~\ref{teor:main},  whereas the
fundamental ingredients of the proof are separately given in the subsequent
three sections. The first one
(Section~\ref{sez:telchi}) concerns perturbation techniques and deals with the 
formal construction of the adiabatic invariant. The other two sections
have a probabilistic nature: the estimate of the marginal
probability is given in Section~\ref{sez:marginale} together with the
estimate of the norm of the time derivative of the adiabatic invariant.
In Section~\ref{sez:condizionata}, we state
Theorem~\ref{teor:correlazioni_generico} in which
the estimate of the spatial correlations is given, which enables us to give
an estimate on the variance of the adiabatic invariant.  The proof of
Theorem~\ref{teor:correlazioni_generico} requires to  apply a technique due to
Dobrushin and Pechersky (see \cite{do2}), and is reported in
Appendix~\ref{app:dim_correlazioni}. In
Section~\ref{sez:definizione} we discuss how a lower bound on the time
autocorrelation provides information on the relaxation time to
equilibrium. The conclusions follow in 
Section~\ref{sez:conclusione}. Most of the proofs 
of a more technical character are given in two appendices.
 
\section{Stabiliy estimate in the Klein Gordon lattice}\label{sez:stab}
In the literature, as a prototype of several models, the so called
Klein Gordon lattice is studied (see \cite{parisi1}--\cite{cgs}). From
a physical point of view, it mimics a
chain of particles, each free to move about a site of a lattice,
subjected both to an on--site restoring nonlinear force
and to a linear coupling with the nearest neighbours. It can also be seen
as a discretization of the one--dimensional $\Phi^4$ model, which plays
a major role in field theory.

The Hamiltonian of such a system, in suitably rescaled
variables, can be written as $H= H_0+H_1$, in which
\begin{equation}\label{eq:ham}
  H_0\equal \sum_{i=1}^N \omega\left(\frac{p_i^2}2+\frac{q_i^2}2\right)
\quad \mbox{and }H_1\equal\ \eps\sum_{i=1}^{N-1}
  \frac{q_iq_{i+1}}{\omega} +\sum_{i=1}^N\frac{q_i^4}{4\omega^2}\ ,
\end{equation}
where $p=(p_1,\ldots,p_N)$ and $q=(q_1,\ldots,q_N)$ are canonically
conjugated variables in the phase space $\mathcal M$,  and $\eps$ is a positive
parameter, while $\omega$ is defined by $\omega\equal \sqrt{1+2\eps}$.
Since we don't want to face in this paper the problem of small divisors, which
typically arises in perturbation theory, we confine ourselves to
the case of small $\eps $, i.e, of small coupling between the
sites.

We aim at showing that, for small enough $\eps$ and sufficiently large
$\beta$, there exists an adiabatic invariant for $H$ (see
Theorem~\ref{teor:main} below). To come to a precise statement, we
need some preliminaries.

As usual, $\langle X \rangle$ will denote the mean value of a
dynamical variable $X$ with respect to the Gibbs measure $\mu$ relative
to the given Hamiltonian $H$ at a given $\beta$, i.e.,
$$
\langle X \rangle \equal \int_{\mathcal M} X(x) \mu(\dif x)\ .
$$
The $L^2(\mathcal M,\mu)$ norm of $X$ is then $\left\|X\right\|\equal
\sqrt{\langle X^2\rangle}$ and its variance $\sigma^2_X$ is defined
according to $\sigma_X^2\equal  \langle X^2\rangle-\left\langle
X\right \rangle^2$. Finally, we also recall that the
correlation coefficient of two dynamical variables $X$ and $Y$ is
\begin{equation}\label{eq:coeff_correlazione}
\rho_{X,Y}\equal \frac{\langle XY\rangle -\langle X\rangle \langle
  Y\rangle} {\sigma_X \sigma_Y}\ ,
\end{equation}
and that $X$ and $Y$ are said to be \emph{uncorrelated} if
$\rho_{X,Y}=0$.

We can now state our main theorem, in which $[\cdot,\cdot]$
denotes Poisson bracket,
\begin{theorem}[Estimate on the adiabatic invariant]\label{teor:main}
 There exist positive constants $\eps^*$, $\kappa$,
 independent of $N$, such that if
 $\eps<\eps^*$ and $\beta>\eps^{-1}$, then there exists
 a polynomial function  $\bar X$ uncorrelated with $H$ such that
 \begin{equation} \label{eq:tesi}
   \frac {\|\, [\bar X,H]\, \|}{\sigma_{\bar X}} \le 
   \exp\left[ -\left( \frac
   1{\kappa \left(\eps+\beta^{-1}\right)}\right)^{1/4}\right] \equal
   \frac 1{\bar t}\ .
 \end{equation}
\end{theorem}
\blank
\textbf{Remark.}
We require $\bar X$ to be uncorrelated with $H$ in order that our
adiabatic invariant be sufficiently different from the Hamiltonian,
which is obviously a constant of motion.

\blank
Before the proof, we point out immediately that this theorem has two
relevant (and strictly related) consequences on the time evolution of
the dynamical variable $\bar X$. They will make clear in which sense
$\bar t$ at the r.h.s. of (\ref{eq:tesi}) can be seen as a
stability time. The
first consequence (Corollary~\ref{cor:prob}) concerns 
the probability $\mathbf P$ that the value of the variable $\bar X$
changes significantly from its original value. Indeed, it entails that
the probability of such a change is  
practically negligible if $t<\bar t$. The second consequence
(Corollary~\ref{cor:autocorr}) is a lower bound
on the time autocorrelation of $\bar X$. We take here as definition of time
autocorrelation of a dynamical variable the following one:
$$
C_X(t)\equal \rho_{X_t,X}\ ,
$$
where $X_t(x)\equal X(\Phi^t x)$, $\Phi^t$ is the flow generated by
$H$ and $\rho$ the correlation coefficient defined by
(\ref{eq:coeff_correlazione}). We have chosen to rescale 
the usual definition, dividing it by $\sigma^2_X$, because the
variance of $X$  is the
natural scale of its autocorrelation, since $C_X(0)=1$ and the inequality
$|C_X(t)|\le 1$ holds for any $t$.

We report here both results, which follow from Theorem~\ref{teor:main}
and from the simple estimate $\| X_t - X\|^2 \le  t^2  \|
[X,H] \|^2 $. The latter can be found in  the proof of Theorem~1 of paper
\cite{carati} and is however reported here in Section~\ref{sez:definizione} in
order to make that section self contained.
\begin{corollario}\label{cor:prob}
In the hypotheses of Theorem~\ref{teor:main}, for any
$\lambda>0$  one has
$$
\mathbf P\left(\left|\bar X_t -\bar X\right|\ge \lambda\,
\sigma_{\bar X}\right) \le \frac 1{\lambda^2}\left(\frac t{\bar t}
  \right)^2\ .
$$
\end{corollario}
\begin{corollario}\label{cor:autocorr}
In the hypotheses of Theorem~\ref{teor:main}, one has
$$
C_{\bar X}(t) \ge 1-\frac 12 \left(\frac
t{\bar t}\right)^2\ .
$$
\end{corollario}
\blank
\textbf{Remark.} We observe that the notion of
stability time for
dynamical systems is not unambiguously defined. In
Section~\ref{sez:definizione} we will provide 
a definition of ``relaxation time'' in terms of time autocorrelation of
dynamical variables, which seems to us significant from a
physical point of view. With such a definition, Theorem~\ref{teor:main} turns
out to mean that the ``relaxation time'' is exponentially long in the
perturbation parameters.

\section{Scheme of the proof of Theorem~\ref{teor:main}}\label{sez:schema}
First we use a variant of the classical construction scheme of
approximate integrals of motion (see \cite{cherry})
in order to perform the construction of the adiabatic
invariant as a formal power series. Precisely, we use the scheme
developed by Giorgilli and Galgani for a direct construction of integrals of
motion (see \cite{giorgilli} and Section~\ref{sez:telchi} for the actual
implementation). It is well known that the series thus obtained are,
in general, divergent, so that the standard procedure consists in
using as approximate integral of motion a truncation of the
series. Denoting  by $Y_n$ the series truncated at order $2n+2$, it
turns out that it has the form
\begin{equation}\label{eq:intprim}
Y_n\equal H_0+ \sum_{j=1}^n P_j(p,q)\ ,
\end{equation}
where $P_j$ are suitable polynomials. In order to make such a quantity
uncorrelated  with $H$,  it is
convenient to consider $X_n\equal Y_n-H$ instead of $Y_n$ itself.

In order to make the construction rigorous, one has to add rigorous
estimates of the variance $\sigma^2_{X_n}$ of $X_n$, and of the $L^2$
norm of $[X_n,H]$. The first step to get such estimates consists in
controlling the structure of the polynomials $P_j$ (which, in
particular, contain only finite range couplings) and the size of their
coefficients. This is done recursively, by a variant of the technique
of the paper \cite{giorg}, which is implemented in
Section~\ref{sez:telchi} (see Lemma~\ref{lemma:coeff_nostro_caso}). We
emphasize that, at variance with the original paper, we obtain here
estimates independent of the number of degrees of freedom.

Then, due to the structure of the polynomials $P_j$, to get the needed
$L^2$ estimates one has  to compute the 
$L^2$ norm with respect to the Gibbs measure of the monomials
appearing in $P_j$. The key step for this computation consists in
giving an upper bound independent of $N$ to the marginal probabilities
of the Gibbs measure. Such an estimate is obtained by adapting
techniques developed by Bogolyubov and Ruelle (see \cite{bogoljubov} and
\cite{rue}) and is reported in Lemma~\ref{lemma:marginale} of
Section~\ref{sez:marginale}. One thus obtains the following bound
\begin{equation}\label{eq:prima}
 \left\| \dot X_n \right\| \le \sqrt{N}\left(\sqrt 2\beta\right )^{-1}
 \left(n!\right)^4 
\left(\beta^{-1}+\eps\right)^n \kappa_1^n  \ ,
\end{equation}
which is valid for a suitable constant $\kappa_1>0$, provided 
$\eps$ is small enough and $\beta$ large enough (see  
Lemma~\ref{lemma:stima_P_punto} of Section~\ref{sez:marginale}).

We emphasize the presence of the factor $\sqrt N$ and that $\kappa_1$
is independent of $N$. It will be shown that actually the l.h.s. of
(\ref{eq:prima}) is of order $\sqrt N$ even if it is the square root
of a sum of $O(N^2)$ terms. This is due to the fact that most of the
terms have zero mean because the measure is even in $p$ and
furthermore the $p$'s are independent variables.

To get the Theorem, one also needs an estimate  of $\sigma_{X_n}$ from
below. This is obtained in two steps, which are based on the remark that
$\sigma_{X_n}\ge \sigma_{X_1} - \sigma_{\mathcal R}$, where $\mathcal
R\equal X_n-X_1$ is a remainder.

First we compute explicitly $X_1$ and estimate from below
$\sigma_{X_1}$, obtaining a bound proportional to $\sqrt N$. Then,
we estimate from above $\sigma_{\mathcal 
  R}$. Precisely, we use techniques introduced by Dobrushin in papers
  \cite{do1,do2} to show that $\sigma_{\mathcal
  R}$ behaves as $\sqrt N$ (see Lemma~\ref{lemma:stima_P} of
Section~\ref{sez:condizionata}). We remark that this is the analogue
of the law of large numbers. We recall that Dobrushin's
techniques enable us to show that spatial
correlations between variables pertaining to
different lattice sites decrease exponentially with the distance
between the sites, so that the monomials appearing in $P_j$ are
essentially independent, and the variance of $P_n$ is
essentially the sum of the variances of each monomial. This leads
to Lemma~\ref{lemma:stima_correlazione} of
Section~\ref{sez:condizionata}, which shows that, for small enough
$\eps$ and large enough $\beta$, for
$n<\kappa_2^{-1/4}(\eps+\beta^{-1})^{-1/4}$ there holds
\begin{equation}\label{eq:seconda}
\sigma_{X_n}\ge \sqrt N(\eps+\beta^{-1})/(8 \beta)\ ,
\end{equation}
where again $\kappa_2$ is a positive constant.

Then one finds the optimal $n$, call it $\bar n$, such that the ratio
$\|[X_{\bar n},H]\|/\sigma_{X_{\bar n}}$ takes the minimal value. 
Notice that, as $n$ belongs to a bounded domain, the minimum can be attained
at the boundary. The optimization  
is immediately done, once the estimates are given both for the $L^2$ norm 
$\|[X_n,H]\|$ of the time--derivative of the quasi
integral of motion $X_n$, and for its variance $\sigma_{X_n}^2$. Then,
the function $\bar X$ satisfying (\ref{eq:tesi}) of
Theorem~\ref{teor:main} is simply given by $\bar X\equal X_{\bar n}-H\rho_{X_{\bar
     n},H}\,\sigma_{X_{\bar n}}/\sigma_H$. The identity  
     $\sigma^2_{\bar X} =(1-\rho^2_{X_{\bar
     n},H}) \sigma^2_{X_{\bar n}}$, together with the upper bound to $\rho_{X_{\bar
     n},H}$ given by 
Lemma~\ref{lemma:stima_correlazione}, enables us to extend all
conclusions from $X_{\bar n}$ to $\bar X$.

\section{Construction of the adiabatic invariant}\label{sez:telchi}
Following  \cite{giorgilli}, we look for the formal integral of motion
by looking for a sequence of polynomials $\chi=
\left\{\chi_s\right\}_{s\ge 1}$ such that
\begin{equation}\label{eq:formale}
\left[H,T_\chi H_0\right]=0\quad\mbox{at any order,}
\end{equation}
where $T_\chi$ is a linear operator, whose action on a polynomial
function $f$ is formally defined by\footnote{Notice
  that in paper \cite{giorgilli} the $\chi_s$ were required to be
  homogeneous polynomials of degree $s+2$.However, there is no problem in
  considering the present more general case.}
\begin{equation}\label{eq:telchi}
  T_\chi f\equal \sum_{s\ge 0} \left(T_\chi f\right)_s\ , \!\quad\! \mbox{with }
  \left(T_\chi f \right)_0\equal f\ , \!\quad\! \left(T_\chi f\right)_s\equal 
  \sum_{j=1}^s \frac js [\chi_j,\left(T_\chi f\right)_{s-j}]\ .
\end{equation}
Inserting the expansion of $T_{\chi} H_0$ and $H$ in
(\ref{eq:formale}) and equating terms of equal order one gets the
system
\begin{equation}\label{eq:formale1}
\Theta_0=H_0\ ,\quad \Theta_s - L_0\chi_s=\Psi_s\quad\mbox{for }s>0\ ,
\end{equation}
where
\begin{equation}\label{eq:determinazione_Psi}
\begin{split}
  \Psi_1&\equal H_1\ , \\
  \Psi_s&\equal - \sum_{l=1}^{s-1}\frac ls \left[\chi_l, \left(T_\chi
    H_0\right)_{s-l}\right] -
  \sum_{l=1}^{s-1} \left(T_\chi \nucleo_l\right)_{s-l}\quad \mbox{for }s\ge
  2\ ,
\end{split}
\end{equation}
$L_0\equal[H_0,\cdot]$ is the homological operator and
(\ref{eq:formale1}) has to be read as an equation for the unknowns
$\chi_s$, $\Theta_s$, which have to belong, respectively, to the
range and to the kernel of the operator $L_0$. By defining the
projections $\Pi_\mathcal{N}$,  $\Pi_\mathcal{R}$, respectively on the
kernel $\mathcal{N}$ and on the range $\mathcal{R}$ of $L_0$, one thus
determines recursively
\begin{equation}\label{eq:determinazione_chi}
\chi_s=-L_0^{-1}\Pi_\mathcal{R}\Psi_s\ , \quad \nucleo_s=\Pi_\mathcal{N}
\Psi_s\quad \mbox{for } s\ge 1\ .
\end{equation} 
The approximate integral of motion is then obtained by truncating the
sequence $T_\chi H_0$ at a suitable order.

We have to estimate the action of the operator $T_\chi$ on
the class of functions $f(p,q)$  we are interested in, in a norm
which is well suited for our 
problem. Such a norm is defined as follows. Let $\mathcal{H}^{r,i}_s$
denote the class of
monomials\footnote{We adopt here the multi--index notation:
  $k=k_1,\ldots, k_N$ and $l=l_1,\ldots,l_N$
  are vectors of integers, with $|k|=|k_1|+\ldots+|k_2|$. So, $p^kq^l=
  p_1^{k_1}\cdot \ldots \cdot p_N^{k_N}q_1^{l_1}\cdot \ldots \cdot
  q_N^{l_N}$.} $p^kq^l$ of degree $s$, i.e.,  with
$|k|+|l|=s$, which furthermore depend on sites that are at most $r$ lattice
steps away from $i$, namely such that $k_j=l_j=0$ if $|i-j|\ge r$.
We denote by $\mathcal{P}_{s,r}$ the set
of all homogeneous polynomials of degree $s$ that can be decomposed as 
\begin{equation}\label{eq:tipo_di_funzioni}
f=\sum_{i=1}^N\sum_{j=1}^{\left|\mathcal{H}^{r,i}_s\right|}
c_{ij}f_{ij}\ ,
\end{equation}
with $f_{ij}\in \mathcal{H}^{r,i}_s$, where
$\left|\mathcal{H}^{r,i}_s\right|$ is the cardinality of
$\mathcal{H}^{r,i}_s$. To $f\in \mathcal P_{s,r}$ we associate a
norm,\footnote{One can check that this is indeed a
  norm.} defined by
\begin{equation}\label{eq:norma_coeff}
  \left\|f\right\|_+\equal  \min\left\{ \max_{i\in\{1,\ldots,N\}} \sum_{j=1}^{
    \left|\mathcal{H}^{r,i}_s\right|} |c_{ij}|\right\} \ ,
\end{equation}
where the minimum is taken over all possible decompositions of $f$.

Now, we can estimate the action of $T_\chi$ on any function $f\in
\mathcal{P}_{s,r}$ according to the following Lemma, which is proved
in Appendix~\ref{app:coeff}.
\begin{lemma}\label{lemma:coeff}
Let $T_\chi$ be the operator defined by (\ref{eq:telchi}), relative
to the sequence $\chi=\{\chi_s\}_{s \ge 0}$ which solves the system of
equations (\ref{eq:determinazione_chi}--\ref{eq:determinazione_Psi})
for the Hamiltonian (\ref{eq:ham}). Then, for any $f(p,q)\in
\mathcal{P}_{2s+2,r}$, one has
$(T_\chi f)_n= \sum_{l=0}^n f_n^{(s+l)} ,
$
where $f_n^{(s+l)}\in \mathcal{P}_{2s+2l+2,r+n-l}$ and 
\begin{equation}\label{eq:induzione_T_chi_2}
\left\|f_n^{(s+l)}\right\|_+\le 2^{6n} 2^{5(n-1)}2^{2s+l+2}
n!\frac{(n+r)!}{r!}\frac{(n+s)!}{s!}
\frac{n!}{l!(n-l)!}\eps^{n-l}\left\|f\right\|_+\ .
\end{equation}
\end{lemma}

Lemma~\ref{lemma:coeff_nostro_caso} below will give bounds to the
adiabatic invariant obtained by truncating at a finite order the
formal power series which defines $T_\chi H_0$. In particular, the
adiabatic invariant will  simply be  
$Y_n= \sum_{s=0}^n \left( T_\chi H_0 \right)_s ,$
so that  the polynomials $P_j$ appearing at the r.h.s. of (\ref{eq:intprim}) of
Theorem~\ref{teor:main} are
\begin{equation}\label{eq:definizione_P_n}
   P_j \equal  \left(T_\chi H_0\right)_j
 \ ,
\end{equation}
while the quantity we will focus on will be
\begin{equation}\label{eq:definizione_X_n}
X_n\equal Y_n-H=-\nucleo_1+\sum_{j=2}^n P_j\ .
\end{equation}
The time derivative of $X_n$ is then given by
\begin{equation}\label{eq:derivata_troncata}
\dot X_n\equal \left[X_n,H\right]=
\left[P_n,H_1\right]\ ,
\end{equation}
which is a polynomial of order $2n+4$. In order to obtain the
estimates of the $L^2$--norm, eventually, it is of interest to 
take into account the parity properties of the operator 
$T_\chi$, with respect to the canonical coordinate $p$. So we define
as $\mathcal{P}^+$ the space of polynomials  of even order
in $p$, and $\mathcal{P}^-$ the space of those of odd order in $p$.

Finally, we can state
\begin{lemma}\label{lemma:coeff_nostro_caso}
For the adiabatic invariant constructed through $T_\chi H_0$ (see
(\ref{eq:definizione_P_n})) one can write
\begin{equation}\label{eq:scomposizione}
  P_n=\sum_{l=0}^n\frac{n!}{l!(n-l)!}\eps^{n-l} P_n^{(l)}\ ,
\end{equation}
where $ P_n^{(l)}\in\mathcal {P}^+ \cap\mathcal{P}_{2l+2,n-l}$ and 
\begin{equation}\label{eq:norma_coeff_P}
\left\|P_n^{(l)}\right\|_+\le \mathcal{D}_n\ ,\quad \mbox{with }
\mathcal{D}_n\equal 2^{12n} 
\left(n!\right)^3\ .
\end{equation}
Furthermore, one has
$$
\left[X_n,H\right]=\sum_{l=0}^{n+1}\frac{(n+1)!}{l!(n+1-l)!}\eps^{n+1-l} \dot
X_n^{(l)}\ ,
$$
with $\dot X_n^{(l)}\in\mathcal{P}^-\cap \mathcal{P}_{2l+2,n+1-l}$ and
\begin{equation}\label{eq:norma_coeff_P_punto}
\left\|\dot X_n^{(l)}\right\|_+\le \mathcal{C}_n\ , \quad\mbox{with }
\mathcal{C}_n \equal 48\cdot 2^{12n}
n!\left((n+1)!\right)^2 \ .
\end{equation}
\end{lemma}
\blank
\textbf{Proof.} The proof of the upper bounds is mainly based on the
  application of   Lemma~\ref{lemma:coeff} to the function
  $H_0\in\mathcal P_{2,0}$,   together with the simple bound
  $\left\| H_0\right\|_+=   \omega\le 2$, which holds for
  small enough  $\eps$. This proves equations (\ref{eq:scomposizione}),
  (\ref{eq:norma_coeff_P}). Then, we use the fact that
  $[X_n,H]=[P_n,H_1]$ and the upper bound to the norm of the Poisson
  brackets of two variables provided by Lemma~\ref{lemma:par_Poisson} of
  Appendix~\ref{app:coeff}. This gives equation
  (\ref{eq:norma_coeff_P_punto}).  

The parity properties are obtained by observing that $[\mathcal
  P^\pm,\mathcal P^\pm]\subset \mathcal P^+$ and $[\mathcal
  P^\pm,\mathcal P^\mp]\subset \mathcal P^-$, as well as
$\Pi_\mathcal{N}(\mathcal{P}^+)\subset \mathcal{P}^+$ and
$\Pi_\mathcal{N}(\mathcal{P}^-)\subset\mathcal{P}^-$ and that the similar
inclusions regarding $\Pi_\mathcal{R}$hold, and then working recursively.
\begin{flushright}Q.E.D.\end{flushright}\blank

\section{Marginal probability estimates}\label{sez:marginale}
The aim of this section is to prove the bound on the norm of $\dot
X_n$ given by the following 
\begin{lemma}\label{lemma:stima_P_punto}
There exist constants $\bar\beta>0$, $\bar\eps>0$, $\kappa_1>0$ such that,
for any $\beta>\bar\beta$ and for any $\eps<\bar\eps$, 
for $\dot X_n$ defined  by (\ref{eq:derivata_troncata}) of
Section~\ref{sez:telchi} one has
\begin{equation}\label{eq:stima_P_punto}
 \left\| \dot X_n \right\| \le \sqrt{N}\left(\sqrt 2\beta\right )^{-1}
 \left(n!\right)^4 
\left(\beta^{-1}+\eps\right)^n \kappa_1^n  \ .
\end{equation}
\end{lemma}
The key tool of the proof is an estimate of the probability
that the coordinates of a finite number $s$ of sites are near some
fixed values. Such an estimate is given in the following
Subection~\ref{sottosez:stima_marginale}, whereas the proof of
Lemma~\ref{lemma:stima_P_punto} is given in
Subection~\ref{sottosez:stima_X_punto}.

\subsection{Estimates on the marginal
  probability}\label{sottosez:stima_marginale}
Everything is trivial for the $p$ coordinates, for which
the measure can be decomposed as a product: from a probabilistic point
of view, this means that every $p_j$ is independent of the $q$ and
of any $p_i$, for $i\neq j$. We focus, instead, on the $q$
coordinates, which are independent of the $p$, but depend on each
other. Then, we must study the relevant part of the density, which is
given by
\begin{equation}\label{eq:definizione_D_N}
  D_N(q_1,\ldots,q_N)\equal \frac{1}{Z_N}\exp\left[-\beta U_N(
      q_1,\ldots,q_N)\right]\ ,
\end{equation}
where $Z_N$ is the ``spatial'' partition function
\begin{equation}\label{eq:definizione_Z_N}
  Z_N\equal \int_{-\infty}^{+\infty}\dif q_1\ldots \int_{-\infty}^{+\infty}\dif q_N
  \,\exp\left[-\beta U_N(q_1,\ldots, q_N)\right]
\end{equation}
and $U_N$ the part of Hamiltonian (\ref{eq:ham}) which depends on
$q$, namely, the potential
$$
U_N\left(q_1\ldots,q_N\right)\equal \sum_{i=1}^N\left(\omega\frac{
q_i^2}{2}+\frac{q_i^4}{4\omega^2}\right) +
\eps \sum_{i=1}^{N-1}\frac{ q_iq_{i+1}}{\omega}\ .
$$

The main point is then to estimate the marginal
probability $F^{(N)}_{s,\xgot}(q_{i_1},\ldots, q_{i_s})$ that we are
going to define. Given a set of indices $i_1<i_2<\ldots<i_s$ we say
that they form a connected block if $i_{j+1}=i_j+1$, i.e., if they
label a ``connected'' chain. We say that a sequence of indices
$i_1<i_2<\ldots<i_s$ form $\xgot$ blocks if the set $\{i_j\}_{j=1}^s$
can be decomposed into $\xgot$ connected blocks, which furthermore are
not connected to each other. Given a set of indices $i_1<i_2<\ldots<i_s$
we define
\begin{equation}\label{eq:definizione_F_N_s}
  F^{(N)}_{s,\xgot}(q_{i_1},\ldots, q_{i_s})\equal  \int_{-\infty}^{+\infty}\dif 
  q_{i_{s+1}}\ldots \int_{-\infty}^{+\infty}\dif  q_{i_N}\, D_N( q_1,\ldots,
  q_N)\ ,
\end{equation}
where $\xgot$ is the number of blocks in the set $\{i_j\}_{j=1}^s$.
We remark here that such a quantity depends on the number of
particles, $N$, but we will find for it an upper bound independent of
$N$. 
In fact, the estimate will depend only on $s$ and $\xgot$, but
not on the precise choice of the sites.

Define the two functions
\begin{equation}\label{eq:def_n_s}
\begin{split}
n_{s,\xgot}(q_{i_1},\ldots,q_{i_s}) &\!\equal\!
\exp\!\left[\!-\beta\!\!\left(\! \sum_{k=1}^s \!
  \left( \frac{q_{i_k}^2}{2\omega} + \frac{q_{i_k}^4}{4
    \omega^2}\right)\!+\!  \eps\!\!\sum_{k,l=1}^s\!
  \delta_{i_l,i_k+1}\!\frac{(q_{i_k}\!\!-q_{i_l})^2}{2\omega}\!\right)\!\right]\\
&\le \exp\!\left(-\beta \sum_{k=1}^s\frac{q_{i_k}^2}{2\omega}\right) \ ,
\end{split}
\end{equation}
\begin{equation}\label{eq:def_n_tilde_s}
\tilde{n}_{s,\xgot}(q_{i_1},\ldots,q_{i_s}) \!\equal\!
\exp\!\left[\!-\beta\!\left( \sum_{k=1}^s 
  \left( \frac{\omega q_{i_k}^2}{2} + \frac{q_{i_k}^4}{4
    \omega^2}\right)\!+  \eps\!\!\sum_{k,l=1}^s
  \delta_{i_l,i_k+1}\frac{q_{i_k}q_{i_l}}{\omega}\!\right) \!\right]\ ,
\end{equation}
where $\delta_{i,j}$ is the Kr\"onecker delta.
\blank
\textbf{Remark.}
Notice that $n_{s,\xgot}$ is the configurational part of the Gibbs
measure of the system with 
variables $q_{i_1},\ldots,q_{i_s}$ and free boundary conditions, apart
from the absence of the normalization factor (i.e., the partition function),
whereas $\tilde{n}_{s,\xgot}$ is the analogous quantity for the same
system, but with fixed boundary conditions.
Thus, they differ only because of the different  
dependence on the coordinates  at the sites lying
on the boundary of the blocks, the number of which,
$\gamma$, satisfies $\xgot\le\gamma\le2\xgot$.
If we denote
by $m_1,\ldots,m_{\gamma}$ the indices of these sites, we can write
the identity  
\begin{equation}\label{eq:rapporto_n_n_tilde}
\frac{n_{s,\xgot}(q_{i_1},\ldots,q_{i_s})}{\tilde{n}_{s,\xgot}(q_{i_1},\ldots,q_{i_s})}
= \prod_{j=1}^\gamma\exp \left(\frac{\beta\eps}{\omega} \alpha_{m_j} q_{m_j}^2
\right)\le \prod_{j=1}^\gamma\exp \left(\frac{\beta\eps}{\omega} q_{m_j}^2
\right)\ ,
\end{equation}
where the factor $\alpha_{m_j}$ is equal to 1 or $1/2$ according to
whether the site $m_j$ is isolated (i.e., the block is composed
of only that site) or not.
\blank

Then the following lemma, which is the main result of the present
subsection, holds
\begin{lemma}\label{lemma:marginale}
There exist constants $\bar\beta>0$, $\bar\eps>0$, $K>0$ and a sequence
$\mathfrak{C}_\xgot >0$ such that, for any $\beta >\bar\beta$ and for any
$\eps<\bar\eps$, one has the inequalities
\begin{equation}\label{eq:maggiorazione_F_N_s}
  F^{(N)}_{s,\xgot}(q_{i_1},\ldots,q_{i_s}) \le \mathfrak{C}_\xgot K^s
\left( \frac{\beta}{2
    \pi\omega}\right)^{s/2}n_{s,\xgot}(q_{i_1},\ldots,q_{i_s})
\end{equation}
and
\begin{equation}\label{eq:minorazione_F_N_s}
  F^{(N)}_{s,\xgot}(q_{i_1}\!,\ldots,\!q_{i_s}\!) \!\ge\!
  \frac{1}{\mathfrak{C}_\xgot} \!\left(\!\! \frac{\beta}{2
    \pi\omega}\!\!\right)^{s/2}\!\!\!\tilde{n}_{s,\xgot}(q_{i_1}\!,\ldots,\!q_{i_s}\!)
  \exp  \!\!\left(
  \!\!-8\eps\xgot\sqrt
  \frac{\beta}{2\omega}\sum_{j=1}^\gamma \!\left|q_{m_j}
  \right|\!\!\right)\ .
\end{equation}
\end{lemma}

The proof of such a lemma
is based on the techniques of paper \cite{bogoljubov},
which apply quite simply to the case of periodic
boundary conditions (see Lemma~\ref{lemma:periodico} below), on
account of the translational invariance. Thus, it is also useful to
introduce the density $\tilde{D}_N$ relative to the periodic system, defined by
\begin{equation}\label{eq:definizione_D_tilde_N}
\tilde{D}_N(q_1,\ldots,q_N)\equal \frac 1{Q_N} \exp\left[-\beta
  U_N(q_1,\ldots,q_N)+\beta \eps q_1q_N\right]\ .
\end{equation}
In this definition there appears the partition function for the
periodic case
\begin{equation}\label{eq:definizione_Q_N}
  Q_N\equal \int_{-\infty}^{+\infty}\dif q_1\ldots \int_{-\infty}^{+\infty}\dif q_N
  \,\exp\left[-\beta U_N(q_1,\ldots, q_N)+\beta \eps
  q_1 q_N\right]\ . 
\end{equation}

For the periodic system it is simple to estimate two relevant
quantities. The former is the ratio
between the partition function for $N-1$ particles and that for $N$
particles, i.e., the ratio $Q_{N-1}/Q_N$. The relation between $Q_N$
and $Z_N$ is then obtained as a particular
case of Lemma~\ref{lemma:rapporto_Z_Q}, which will be stated
later on. 
The latter is the probability, evaluated with respect to the
density for $N$ particles, that the coordinates of $r$
particles have an absolute value smaller than
$\Theta\sqrt{2\omega/\beta}$, for a given $\Theta$. 
In other terms, we need an estimate of the following quantity
\begin{equation}\label{definizione_P_N}
\begin{split}
 \mbox{\bf{P}}_N\!\!\left(|q_1|<\Theta\sqrt{\frac{2\omega}{\beta}}\wedge
 \ldots\wedge
 |q_r|<\right.& \left.\Theta\sqrt{\frac{2\omega}{\beta}} \right) \equal 
 \int_{-\infty}^{+\infty}\!\!\!\dif
 q_1\ldots\! \int_{-\infty}^{+\infty}\!\!\!\dif q_N
 {\bf{1}}_{|q_1|<\Theta\sqrt{2\omega/\beta}} \\
& \times\ldots \times{\bf{1}}_{|q_r|<\Theta\sqrt{2\omega/\beta}}\, \tilde{D}_N(q_1,
 \ldots,q_N) \ ,
\end{split}
\end{equation}
in which ${\bf 1}_A$ is the indicator function of the set $A$. We can
now give the mentioned estimates by the following lemma, whose proof
is deferred to Appendix~\ref{app:dim_lemma_periodico}.
\begin{lemma}\label{lemma:periodico}
There exist constants $\beta_0>0$, $\eps_0>0$, $K_0>2$ such that, for any
  $\beta >\beta_0$ and $\eps<\eps_0$, one has
\begin{equation}\label{eq:maggiorazione_frazione_Q_N}
\frac{Q_{N-1}}{Q_N}\le K_0\sqrt{\frac{\beta}{2\pi\omega}}\ .
\end{equation}
Furthermore, if $\Theta\ge2\sqrt{r\log (4r K_0)}$, one has
\begin{equation}\label{eq:minorazione_P_N}
\mbox{\bf{P}}_N\left(|q_1|<\Theta\sqrt{\frac{2\omega}\beta}\wedge \ldots\wedge
 |q_r|<\Theta\sqrt{\frac{2\omega}\beta} \right)\ge \frac 12\ .
\end{equation}
\end{lemma}

This result enables us to give the proof of
Lemma~\ref{lemma:marginale}.

\vskip 1em
\noindent
\blank \textbf{Proof of Lemma~\ref{lemma:marginale}}
First write
$$
D_N\left( q_1.\ldots,q_N\right)=n_{N-s,\xgot'}\left(q_{i_{s+1}},\ldots ,
  q_{i_N}) \right)  n_{s,\xgot}\left(q_{i_1},\ldots,q_{i_s}\right)
  I\left(q_1, \ldots,q_N\right)\ ,
$$
for a suitable $\xgot'$, with $\xgot-1\le \xgot'\le \xgot+1$ (the lower and
the upper bound are attained, respectively, if both $1$ and $N$ are
contained in $i_1,\ldots,i_s$ or none of them), where $I$ contains the
terms of interaction between the ``internal'' and the external part of
the system. Remarking that $I\le1$, one gets
\begin{eqnarray}\label{prima_maggiorazione}
  F^{(N)}_{s,\xgot}(q_{i_1},\ldots,q_{i_s})&\le&
  \frac{1}{Z_N}\left(
   \int_{-\infty}^{+\infty}\dif q_{i_{s+1}} \ldots
  \int_{-\infty}^{+\infty}\dif  q_{i_N}\times\right. \nonumber\\
&&\left.\frac{}{}\times n_{N-s,\xgot'}(q_{i_{s+1}},\ldots ,
  q_{i_N}) \right)  n_{s,\xgot}(q_{i_1},\ldots,q_{i_s})\ .
\end{eqnarray}

We now have to estimate the integral appearing in
(\ref{prima_maggiorazione}). More in general, in the course
of the proof we  need to estimate  integrals of a
similar type. This will be done in Lemma~\ref{lemma:rapporto_Z_Q},
which will be given in a while. Introduce the quantities:
\begin{equation}\label{eq:definizione_Q_corsivo}
\begin{split}
\bar{\mathcal{Q}}^\xgot_M &\equal  \inf_{B(M,\xgot)}\int_{-\infty}^{+\infty}\dif q_1 \ldots
  \int_{-\infty}^{+\infty}\dif  q_M n_{M,\xgot}(q_1,\ldots,q_M)\ ,\\
\mathcal{Q}^\xgot_M &\equal  \sup_{B(M,\xgot)}\int_{-\infty}^{+\infty}\dif q_1 \ldots
  \int_{-\infty}^{+\infty}\dif  q_M n_{M,\xgot}(q_1,\ldots,q_M)\ ,
\end{split}
\end{equation}
where $B(M,\xgot)$ denotes
the collection
of all possible partitions of
$M$ indices in $\xgot$ blocks.
It is also convenient to consider the quantities defined
in a similar way, by integrating $\tilde{n}_{M,\xgot}$ in place of
$n_{M,\xgot}$, namely
\begin{equation}\label{eq:definizione_Z}
\begin{split}
\bar{Z}^\xgot_M &\equal  \inf_{B(M,\xgot)}\int_{-\infty}^{+\infty}\dif q_1 \ldots
  \int_{-\infty}^{+\infty}\dif  q_M \tilde{n}_{M,\xgot}(q_1,\ldots,q_M)\ ,\\
Z^\xgot_M &\equal  \sup_{B(M,\xgot)}\int_{-\infty}^{+\infty}\dif q_1 \ldots
  \int_{-\infty}^{+\infty}\dif  q_M \tilde{n}_{M,\xgot}(q_1,\ldots,q_M)\ .
\end{split}
\end{equation}
It is easily shown that for $\xgot=1$ one has $\bar{Z}^1_M=Z^1_M=Z_N$. 
In order to link them to $Q_N$ and to each
other, we use the following lemma, the proof of which is deferred to
Appendix~\ref{app:dim_lemma_rapporto}.
\begin{lemma}\label{lemma:rapporto_Z_Q}
Let $\beta_0>0$, $\eps_0>0$ and $K_0>2$ be constants such that
Lemma~\ref{lemma:periodico} holds. Then, for any $\beta
>\beta_0$ and any $\eps<\eps_0$, the inequalities
\begin{equation}\label{eq:minorazione_rapporto_Z_Q}
\frac{\bar{\mathcal{Q}}^\xgot_M}{Q_M}\ge1\ ,\quad\frac{\bar Z^\xgot_M}{Q_M}\ge \frac
12 \left(8\xgot K_0\right)^{-32\eps_0 \xgot^2}
\end{equation}
hold. Furthermore, the chain of inequalities
\begin{equation}\label{eq:maggiorazione_rapporto_Z_Q}
\frac{Z^\xgot_M}{Q_M}\le\frac{\mathcal{Q}^\xgot_M}{Q_M}\le 2 K_0^\xgot
\exp\left(4\xgot\eps_0\bar\kappa(\xgot,K_0)\right) \ ,
\end{equation}
holds, where $\bar \kappa(\xgot,K_0)$ is the solution of the equation
\begin{equation}\label{eq:determinazione_kappa_barra}
K_0^{2\xgot} \Gamma(\xgot,\bar \kappa)=\frac 12\ ,
\end{equation}
$\Gamma(s,x)$ being the upper regularized Gamma
function
\begin{equation}\label{eq:gamma_reg}
\Gamma(s,x) \equal  \frac 1{(s-1)!}\int_{x}^{+\infty} t^{s-1} e^{-t} \dif t \ .
\end{equation}
\end{lemma}
The previous lemma enables one to see that
$Z_N^{-1}\le 2(8K_0)^{32\eps_0}/Q_N,
$
while the integral appearing in (\ref{prima_maggiorazione}) is estimated
by
$$
\mathcal{Q}^{\xgot'}_{N-s}\le
2K_0^{\xgot+1}\exp\left[4(\xgot+1)\eps_0\bar\kappa(\xgot+1,K_0)\right] Q_{N-s}\ .
$$
Thus, due to relation
(\ref{eq:maggiorazione_frazione_Q_N}) of Lemma~\ref{lemma:periodico},
one easily sees that
\begin{equation}\label{maggiorazione_periodico}
\frac{Q_{N-s}}{Q_N}= \prod_{i=1}^s\frac{Q_{N-i}}{Q_{N-i+1}}\le
K_0^s\left(\frac{\beta}{2\pi\omega}\right)^{s/2}\ ,
\end{equation}
so that (\ref{eq:maggiorazione_F_N_s}) is proved, taking
\begin{equation}\label{eq:prima_maggiorazione_gotico}
\mathfrak{C}_\xgot \ge
2^{96\eps_0+2}K_0^{\xgot+1+32\eps_0}\exp\left[4(\xgot+1)\eps_0 \bar\kappa
  (\xgot+1,K_0)\right]\ .
\end{equation}

We come now to the proof of (\ref{eq:minorazione_F_N_s}). To this end,
we write
\begin{equation}\nonumber
\begin{split}
D_N(q_1,\ldots,q_N)&=\frac {Q_{N-s}}{Z_N} \tilde{D}_{N-s}(q_{i_{s+1}},\ldots,q_{i_N}) \tilde{n}_{s,\xgot}(q_{i_1},\ldots,
q_{i_s}) 
\\
&\times G(q_{m_1},\ldots,q_{m_\gamma},q_{l_1},\ldots,q_{{l_{\gamma'}}}) 
\ ,
\end{split}
\end{equation}
where the sites $l_i$ are the ones which are contiguous to the blocks,
but not contained in them, taken by keeping the relative
order. Furthermore, due to the periodicity there appear factors
depending on $q_1$ and $q_N$, if the sites 1 and $N$ are not contained
in $i_1,\ldots,i_s$. In this case, we
put $q_{l_1}=q_1$ and $q_{l_{\gamma'}}=q_N$. We denote by $\gamma'$, with
$\gamma'\le2\xgot+2$, the number of such indices. 
The explicit expression of the function $G$ is complicated. Plainly,
it represents the product of the factors $\exp(-\sum_j\beta\eps 
q_{m_j}q_{l_i}/\omega)$  among all sites $l_i$ contiguous to
$m_j$, and just the factor $\exp(\beta\eps 
q_{l_i}q_{l_{i+1}}/\omega)$, when $l_i$ and $l_{i+1}$ belong to
different blocks.\footnote{Notice that the index $i$ lies in
  $\{1,\ldots,\gamma'\}$. But if in the expression $\exp(\beta\eps
  q_{l_i}q_{l_{i+1}}/\omega)$ there appears $q_{l_{\gamma'+1}}$ then one has to
  intend simply $l_{\gamma'+1}$ as $l_1$.}   
In any case, a lower bound to $G$ in the region 
$$
\mathcal{A} \equal  \left\{|q_{l_1}|\!<\!2\xgot \sqrt{\log(4 K_0)}
\sqrt{2\omega /\beta} 
\;\wedge\;\ldots\;\wedge\; |q_{l_{\gamma'}}|\!<\! 2\xgot \sqrt{\log(4 K_0)}
\sqrt{2\omega /\beta}
\right\}$$
is given by
$$
  G \ge
 \exp\left(  -4\eps\xgot\sqrt \frac{\beta}{2\omega}\sqrt{\log(4
   K_0)}\sum_{j=1}^\gamma \left|q_{m_j}
  \right|\right) (4K_0)^{-8\eps (\xgot^2+\xgot)} \ .
$$
So, we can write 
\begin{equation}\label{eq:minorazione_intermedia_F_N_s}
\begin{split}
 F^{(N)}_{s,\xgot}(q_{i_1},\ldots,q_{i_s}) &\ge
  \tilde{n}_s(q_{i_1},\ldots,q_{i_s})(4K_0)^{-8\eps (\xgot^2+\xgot)}\frac{Q_{N-s}}{Z_N} \,
  \mbox{\bf{P}}_{N-s}\left(\mathcal{A}\right)  \\
&\times  \exp\left(  -4\eps\xgot\sqrt
  \frac{\beta}{2\omega}\sqrt{\log(4 K_0)}\sum_{j=1}^\gamma \left|q_{m_j}
  \right|\right)\ ,
\end{split}
\end{equation}
where the (positive) contribution of the integral over
$\mathcal A^c$ was neglected.

The term with $\mbox{\bf{P}}_{N-s}(\mathcal A)$ in
(\ref{eq:minorazione_intermedia_F_N_s}) is bounded from below by
relation (\ref{eq:minorazione_P_N}) of Lemma~\ref{lemma:periodico}. As
for the fraction, by Lemma~\ref{lemma:rapporto_Z_Q}
we obtain
$$
Q_{N-s}\ge \frac{1}{2K_0\exp(8\eps_0\bar\kappa(1,K_0))}\mathcal{Q}_{N-s}^1\ .
$$
Now, operating as in the deduction of formula
(\ref{maggiorazione_periodico}), it is sufficient to observe that
$\mathcal{Q}_{N-1}^1\ge \sqrt{\beta/(2\pi\omega)} \mathcal{Q}_N^1
$
to obtain $\mathcal{Q}_{N-s}^1\ge \left(\beta/2(\pi\omega)\right)^{s/2}
\mathcal{Q}_N^1$. Then, choosing $\bar \eps$, $\bar \beta$ such that
$K_0\le e^4/4$ and observing that $ \mathcal{Q}_N^1\ge Z_N$, one gets
(\ref{eq:minorazione_F_N_s}) with
\begin{equation}\label{eq:seconda_maggiorazione_gotico}
\mathfrak{C}_\xgot \ge
\left(4K_0\right)^{8\eps_0(\xgot^2+\xgot)+1}\exp(8\eps_0\bar\kappa(1,K_0)
\ .
\end{equation}
Finally, $\mathfrak{C}_\xgot$ can be chosen as the maximum of the
r.h.s. of (\ref{eq:prima_maggiorazione_gotico}) and of
(\ref{eq:seconda_maggiorazione_gotico}) . This concludes the proof.
\begin{flushright}Q.E.D.\end{flushright}\blank

\subsection{Estimate of $\|\dot X_n\|$}\label{sottosez:stima_X_punto}
We apply directly inequality (\ref{eq:maggiorazione_F_N_s}) to get
the proof of Lemma~\ref{lemma:stima_P_punto}, using the fact that
such a quantity is a sum of polynomials depending at most on $2n+3$
sites, as can be seen by Lemma~\ref{lemma:coeff_nostro_caso} of
Section~\ref{sez:telchi}.

\blank \textbf{Proof of Lemma~\ref{lemma:stima_P_punto}} The key ingredient
  of the proof is, as 
stated in Section~\ref{sez:telchi}, that the polynomials $P_n$ are even in
the $p$ coordinates, so that the $\dot X_n$'s are odd in the
$p$. On account of that, $\dot X_n^2$ is a sum in which the terms
coming from the product of two monomials depending on separated groups
of sites contain at least one $p_i$ to an odd power. Since the measure
is even with respect to any $p$, these terms have a vanishing integral.

We formalize this way of reasoning by decomposing $\dot X_n$ as
$
\dot X_n= \sum_{i=1}^N f_i,
$
where the $f_i$'s are polynomials depending at most on the sites
between $i-n-1$ and $i+n+1$. Then, the $L^2$--norm of $\dot X_n$ is
expressed according to
$
\left\|\dot X_n\right\|^2=\sum_{i,j=1}^N \langle f_i f_j\rangle\ .
$
In this sum, all the terms with $|i-j|>2n+2$ vanish, while the other ones
are estimated in terms of $\left\|\dot X_n\right\|_+$ in the following
way.

On account of Lemma~\ref{lemma:coeff_nostro_caso}, we can write
$$
f_i=\sum_{l=0}^{n+1}\frac{(n+1)!}{l!(n+1-l)!}\eps^{n+1-l}\sum_{s=1}^{
  |\mathcal{H}^{n+1-l,i}_{2l+2}|}
c_{is,l} f_{is}^{(l)}\ ,
$$
in which $f_{is}^{(l)}$ is a monomial in
$\mathcal{H}^{n+1-l,i}_{2l+2}$ and the decomposition in these
monomials is performed in such a way that
$\sup_{i,l}\sum_s |c_{is,l}|\le \mathcal C_n$.
Then, we sum on $j$ and obtain that
$$
\left\|\dot X_n\right\|^2\le(4n+5)\, \mathcal
C_n^2\sum_{i=1}^N\sum_{l=0}^{n+1}\frac{(2n+2)!}{l!(2n+2-
  l)!}\eps^{2n+2- l}\sup_{g\in\mathcal{H}^{2n+2-l,i}_{2l+4}}
\langle g\rangle\ ,
$$
where we used the fact that the only nonvanishing
contributions to the integral come from the product of $f_{js}^{(l-r)}\in
\mathcal{H}^{n+1-l+r,j}_{2l-2r+2}$ and $f_{km}^{(r)}\in
\mathcal{H}^{n+1-r,k}_{2r+2}$, for $|j-k|\le 2n+2-l$, so that $g\equal
f_{js}^{(l-r)} f_{km}^{(r)}\in \mathcal{H}^{2n+2-l,i}_{2l+4}$, for a
suitable $i$ between $j$ and $k$.

Then, we make use of (\ref{eq:maggiorazione_F_N_s}) together with the
estimate (\ref{eq:def_n_s}) for $n_{s,\xgot}$ to bound the mean value of
any function in
$\mathcal{H}^{2n+2-l,i}_{2l+4}$. In fact, one has
\begin{equation*}
\begin{split}
\sup_{g\in\mathcal{H}^{2n+2-l,i}_{2l+4}} \langle g\rangle &\le
\mathfrak{C}_1K^{4n-2l+5}\sqrt\frac{\beta}{2\pi\omega} 
\int_{-\infty}^{\infty} x^{2l+4} \exp\left(-\frac{\beta
}{2\omega} x^2\right)\dif x\\
&= \mathfrak{C}_1 K^{4n-2l+5} \left(
\frac{2\omega}{\beta} \right)^{l+2}  \frac{(2l+3)!!}{2^{l+2}}\ .
\end{split}
\end{equation*}
So, the inequality
$$
\left\|\dot X_n\right\|^2 \le \mathfrak{C}_1K^{4n+5} (4n+5) (2n+4)!\,
(2\omega)^{2n+4} \beta^{-2}\left( \eps+\frac 1{\beta}\right)^{2n+2}N
\mathcal{C}^2_n
$$
holds. Thus, choosing a suitable $\kappa_1>0$ and using the value of
$\mathcal{C}_n$ given by (\ref{eq:norma_coeff_P_punto}) of
Lemma~\ref{lemma:coeff_nostro_caso}, inequality (\ref{eq:stima_P_punto}) is
satisfied.
\begin{flushright}Q.E.D.\end{flushright}\blank

\section{Estimate of the variance of the adiabatic
  invariant}\label{sez:condizionata}
In the present Section we prove the following
Lemma~\ref{lemma:stima_correlazione}, which was used in the proof of
Theorem~\ref{teor:main}. The lemma concerns
 estimates on the variance $\sigma^2_{X_n}$ and on
the correlation $\rho_{X_n,H}$ of the adiabatic invariant and reads 
\begin{lemma}\label{lemma:stima_correlazione}
 There exist positive constants  $\tilde{\eps}>0$, $\kappa_2>0$,
 $\kappa_3>1$, such that, for any $\eps<\tilde \eps$, for any
 $\beta>\eps^{-1}$ and for
 $n<\kappa_2^{-1/4}(\eps+\beta^{-1})^{-1/4}$, with $X_n$ defined by
 (\ref{eq:definizione_X_n}), the following inequalities hold:
\begin{equation}\label{eq:minorazione_sigma_X_n}
\sigma_{X_n}\ge \sqrt N\, \frac{\eps+\beta^{-1}}{8 \beta}
\end{equation}
and
\begin{equation}\label{eq:stima_correlazione}
\left|\rho_{X_n,H}\right|\le
\left(1+\frac 1{\kappa_3}\frac{\eps^2}{\left(\eps+ \beta^{-1}
  \right)^2} \right)^{-1/2}\ .
\end{equation}
\end{lemma}

The proof of this lemma requires the study the spatial correlations
between quantities depending on two separate blocks. The study of
these properties has to be performed within the general frame of
Gibbsian fields and conditional probabilities. In the present Section
we provide the necessary 
notions and give a proposition of a general character concerning the
decay of spatial correlations for lattices with finite
range of interaction, i.e, Theorem~\ref{teor:correlazioni_generico} of
Subsection~\ref{sottosez:corrspaziali}, from which it will be
possible to finally come to the proof of
Lemma~\ref{lemma:stima_correlazione}, which will be given in
Subsection~\ref{sottosez:stima_sigma_X}.

Our treatment of conditional probabilities is inspired in particular by the
work of Dobrushin (see \cite{do1}). More precisely, we will make
reference to paper \cite{do1} for the main ideas, and to the
subsequent beautiful but underestimated subsequent paper \cite{do2},
by Dobrushin and Pechersky, for a more direct relation to our problem.
As a matter of fact, most of the
ideas of this section are already contained in works \cite{do1} and
\cite{do2}, but the explicit result on the spatial correlations given here
required some additional work. We recall that, since Gibbsian fields and the
related techniques were introduced in order to deal 
with infinite lattices, our result holds even if the
number of sites tends to infinity.

The present section is structured as follows: in
Subsection~\ref{sottosez:corr_e_cond} the link between spatial correlations and
conditional probabilities is shown, and in
Subsection~\ref{sottosez:corrspaziali} we state
Theorem~\ref{teor:correlazioni_generico}, whose proof is deferred to
Appendix~\ref{app:dim_correlazioni}. Such a result is used in
order to obtain an upper bound to $\sigma_{P_n}$, stated in
Lemma~\ref{lemma:stima_P} of Subection~\ref{sottosez:stima_sigma_P},
whence the proof of Lemma~\ref{lemma:stima_correlazione} easily follows,
as shown in Subection~\ref{sottosez:stima_sigma_X}.

\subsection{Link between spatial correlations and conditional
  probability}\label{sottosez:corr_e_cond}
In order to prove Lemma~\ref{lemma:stima_correlazione} we have to estimate
quantities such as
$
\langle fg\rangle-\langle f\rangle \langle g\rangle
$,
relative to the Gibbs measure $\mu$, where $f$ is a function
which depends  on sites belonging to a set 
$\tilde V$, while $g$ depends only on sites in  $V$,
with $V\cap \tilde V =\emptyset$. Our aim is to show that such
correlations decrease as the distance between 
$\tilde V$ and $V$ increases, where the distance $d(V,\tilde V)$
is defined for example as
$d(V,\tilde V)\equal  \inf_{\indmono\in V,\indamono\in \tilde V}
|\indmono- \indamono|$.

We start showing the relation between the
spatial correlations and the conditional probability in a setting more
general than ours. We consider as given a measure $\mu$ on $\mathbb
R^{|\T|}$, with $\T\subset \mathbb Z^\nu$,
which induces on the measurable set $A\subset \mathbb R^{|\tilde V|}$
the probability
$$
P_{\tilde V}(A)\equal \int_{\mathbb R^{|\T|}} \dif \mu(x)
{\bf 1}_{A\times \T\backslash \tilde V}(x)\ ,
$$
where ${\bf 1}_A$ is the indicator function of the set $A$. One can
express the quantity we are interested in as
\begin{equation}\label{eq:correlazioni_condizionato}
\langle fg\rangle\!-\!\langle f\rangle \langle g\rangle\! =\!\! \int_{\mathbb
  R^{|\tilde V|}} \!\!f(\vett x) P_{\tilde V}(\dif \vett x) \!\left(
\!\int_{\mathbb R^{|V|}}\!\!  g(\vett y) P_V (\dif \vett y|\dif \vett x) -
\!\!\int_{\mathbb R^{| V|}}\!\!  g(\vett y) P_V (\dif \vett
y)\right)\ ,
\end{equation}
where $P_V(B|A)$ represents the conditional probability of the
measurable set $B\subset \mathbb R^{|V|}$, once $A$ is given.
So, in order to estimate the
correlation between two functions, it is sufficient to  estimate the
difference enclosed in brackets at the r.h.s. of
(\ref{eq:correlazioni_condizionato}). Now, we notice that for any pair
of probabilities $P$ and $\tilde P$ on $\mathbb R^{|V|}$, one has
\begin{equation}\label{eq:diff_misure_1}
\begin{split}
\left|\int_{\mathbb R^{| V|}}  g(\vett x) P(\dif \vett
x)-\int_{\mathbb R^{| V|}}  g(\vett y) \tilde P(\dif \vett y)\right| \le
\int_{\mathbb R^{|V|}\times \mathbb R^{|V|}} & \left( |g(\vett x)|
+ |g(\vett y)| \right) \\
&\times {\bf 1}_{\vett x\neq \vett y}\, Q(\dif \vett
x,\dif \vett y)\ ,
\end{split}
\end{equation}
in which $Q$
is \emph{any} probability on $\mathbb R^{|V|}\times \mathbb R^{|V|}$
such that  $P$ and $\tilde P$ are its marginal probabilities. In other
terms $Q$ is a joint probability of $P$ and $\tilde P$, i.e.,
for any measurable $B\subset \mathbb R^{|V|}$ one has
\begin{equation}\label{eq:congiunta}
P(B)=Q\left(B\times \mathbb R^{|V|}\right) \quad \mbox{and } \tilde
P(B)=Q \left(\mathbb R^{|V|}\times B\right)\ .
\end{equation} 
Remark here that $Q$ is not unique: indeed, such a probability
provides also a way to define a distance between two probabilities
defined on the same set $V$ of indices, by
\begin{equation}\label{eq:def_distanza_prob}
D(P,\tilde P)\equal \inf_Q \int_{\mathbb R^{|V|}\times\mathbb R^{|V|}}
{\bf 1}_{\vett x\neq\vett y}\, Q(\dif \vett x,\dif \vett y)\ .
\end{equation}
We stress that the infimum is attained, i.e., there exists a
probability measure $\bar Q(\dif \vett x,\dif \vett y)$ such that
$D(P,\tilde P)= \int
{\bf 1}_{\vett x\neq\vett y} \bar Q(\dif \vett x,\dif \vett y)$ (see
Lemma~1 of paper \cite {do2}).
For the following, we suppose that it is possible to find
a compact function $h$,\footnote{See
  later for the definition of a compact function, according to the
  convention of paper \cite{do2}.} with domain
in $\mathbb R$, such that
\begin{equation}\label{eq:maggiorazione_g_h}
|g(\vett x)|\le \sum_{\indmono\in V}h(x_\indmono)\ ,
\end{equation}
as is the case for the monomials we are dealing with. The bounds will then
be given in terms of $h$.
Now, observing that ${\bf 1}_{\vett x\neq \vett y}\le
\sum_{\indmono\in V} {\bf 1}_{x_\indmono\neq y_\indmono}$,  we can
rewrite (\ref{eq:diff_misure_1}) as 
\begin{equation}\label{eq:diff_misure_2}
\begin{split}
\left|\int_{\mathbb R^{| V|}}  g(\vett x) P(\dif \vett
x)-\int_{\mathbb R^{| V|}}  g(\vett y) \tilde P(\dif \vett y)\right|
\le 
\sum_{\indmono,\indamono\in V}\int_{\mathbb R^2} &\left( h(x_\indamono)
+ h(y_\indamono) \right) \\
&\times {\bf 1}_{x_\indmono\neq y_\indmono} Q(\dif\vett
x,\dif\vett y)\ .
\end{split}
\end{equation}
This way, we can make a direct connectino with paper \cite{do2}, in which the
problem of estimating the r.h.s. of the above expression is dealt with. 
We summarize here the results and the methods we need.

\subsection{Main argument and theorem on correlations}\label{sottosez:corrspaziali}

In the quoted work \cite{do2}, the framework is more general than  ours,
because it deals with the problem of defining a ``probability'' for the
configuration of an actually infinite $\nu$-dimensional lattice of
particle, in terms of 
the set of conditional probabilities on each site, 
which is called the \emph{specification} $\Gamma$.
Now, our case is in principle different, because our lattice is
finite, and the probability is defined through the Gibbs measure.
In particular, the specification too is assigned by such a measure. 

However, as proved in  \cite{do2},
under suitable assumptions  assigning the specification uniquely
determines the probability,
i.e. the Gibbsian field, which, in our case, turns out to be precisely
that of Gibbs. So, in our case, it is equivalent to speak in terms of
specification or in terms of measure. Indeed,
in this subsection we will speak in terms of
specifications, and in the following one we will show that the
specification determined  by Gibbs measure (\ref{eq:gibbs}) with the
Hamiltonian (\ref{eq:ham}) satisfies the
assumptions of \cite{do2} (i.e. Conditions~\ref{cond:compattezza} and
\ref{cond:contratt} below).

We  notice that the r.h.s. of (\ref{eq:diff_misure_2}) can be bounded
from above if one estimates the quantity 
\begin{equation}\label{eq:def_lambda}
\lambda(\inda,\ind) \equal  
\max \left\{ \mbox{\bf E} \left[ 
 {\bf 1}_{\xi_\inda^1\neq \xi_\inda^2} h\left(  \xi^1_\ind
 \right)\right], \mbox{\bf E} \left[ 
 {\bf 1}_{\xi_\inda^1\neq \xi_\inda^2} h\left(  \xi^2_\ind \right)\right]
\right\}\ ,
\end{equation}
where $\xi^1$ and $\xi^2$ are two Gibbsian fields which assign,
respectively, the probabilities $P$ and $\tilde P$ appearing in
(\ref{eq:diff_misure_2}) and the
expectations are obtained by integrating over a joint probability
$Q$ of the two fields. Indeed, in \cite{do2} an upper bound just to
$\lambda(\inda,\ind)$ is given, by requiring that two
suitable conditions are satisfied.

So, by adopting the same techniques of \cite{do2}
we bound from above the r.h.s. (\ref{eq:diff_misure_2}) and, thus,
the l.h.s. of (\ref{eq:correlazioni_condizionato}). Such a bound is
given in Theorem~\ref{teor:correlazioni_generico} below. In  order to
state it, we recall the main notations of \cite{do2}.

First, we consider a lattice of sites contained in $\T\subset \mathbb
Z^\nu$, and a finite--range specification with a radius of interaction $r$
(this means that the conditional probability at site $\ind$  does not depend on
the conditioning at sites $\inda$  for $|\ind -\inda|>r$). Then, for a
vector $\vett x\in\mathbb R^{|\T|}$ we denote by $P_{\ind,\vett x}(\dif x)$ 
the probability distribution conditioned to $\vett x$ everywhere but
at site $\ind $. The specification $\Gamma$ is defined by
$$ 
\Gamma \equal \left\{ P_{\ind,\vett x} : \ind\in \T, \vett
     x \in \mathbb{R}^{|T|} \right\} \ .
$$
Furthermore, we will say that a continuous positive
function $h$ on a metric space $\mathfrak{X}$ is compact if, for any $k\ge 0$, the
set $\{x\in \mathfrak{X}: h(x)\le k\}$ is compact.

For a fixed integer $\range$, let
$
\partial_\range V\equal \left\{\inda \in \T: \inda\not \in V,
\min_{\indb\in V}|\inda-\indb|\le
\range\right\}
$
be the boundary of thickness $\range$ of a set $V\subset \T$. We call
$a$ the number of indices such that $|\ind|\le r$, $\ind\neq 0$, where
$\range$ is the range of interaction.

If $Z_0$ is a maximal subgroup of $\mathbb{Z}^\nu$ satisfying the condition
$|\inda-\indb|>\range$, for $\inda,\indb\in Z_0$, we denote by $b$ the
number of elements in the factor group $\mathbb{Z}^\nu\backslash Z_0$.

The conditions of paper \cite{do2} (which are
hypotheses on the specification $\Gamma$, once the compact
function $h$ is given) are the following
\begin{condizione}[Compactness]\label{cond:compattezza}
  Let $h$ be a compact function on $\mathbb{R}$ and let  $C\ge 0$ and
  $c_\inda\ge 0$, for
  $|\inda|\le \range$, $\inda\neq 0$, be some constants. We suppose that
\begin{enumerate}
  \item 
$\displaystyle\delta\equal \sum_{|\inda|\le\range\,,\,\inda\neq 0}c_\inda<
    \frac{1}{a^b}$ ; 
  \item for any $\ind\in \T$ and any $\vett x\in \mathbb R^{|\T|}$ one
    has
$$
\int_{\mathbb R} h(x)P_{\ind,\vett x}(\dif x)\le C+\sum_{\inda\in\partial_\range\{\ind\}}
c_{\inda-\ind} h\left(x_\inda\right) \ .
$$
\end{enumerate}
\end{condizione}
\begin{condizione}[Contractivity]\label{cond:contratt}
  Let $\bar{K}\ge 0$ and $k_\inda=k_\inda(\bar{K})\ge 0$, for
  $|\inda|\le\range$, $\inda\neq 0$, be constants and $h$ be a compact
  function. We   suppose that
\begin{enumerate}
  \item  
$\displaystyle\alpha\equal \sum_{|\inda|\le\range\,,\,\inda\neq 0}k_\inda< 1$ ;
  \item for any $\ind\in \T$ and any pair of configurations $\vett
    x^1,\vett x^2\in \mathbb R^{|\T|}$ such that
$$
\max_{\inda\in \partial_\range\{\ind\}} \max \left\{
h\left(x^1_\inda\right), h\left(x^2_\inda\right) \right\}\le \bar{K}\ ,
$$
one has the inequality
$$
D\left(P_{\ind,\vett x^1},P_{\ind,\vett x^2}\right)\le \sum_{\inda\in\partial_\range\{\ind\}}
k_{\inda-\ind} {\bf 1}_{x^1_\inda\neq x^2_\inda} \ ,
$$
where $D(\cdot,\cdot)$ is the distance defined by
(\ref{eq:def_distanza_prob}).
\end{enumerate}
\end{condizione}

The set of specifications (i.e., the sets of conditional
probabilities on every site) which satisfy
Condition~\ref{cond:compattezza} for the constants $C,\delta$ and the
compact function $h$ will be denoted by $\Theta(h,C,\delta)$. We will
instead denote by $\Delta(h,\bar{K},\alpha)$ the set of
specifications which satisfy Condition~\ref{cond:contratt} for the
constants $\bar{K},\alpha$ and the compact function $h$.

If the specification satisfies Conditions~\ref{cond:compattezza}
and \ref{cond:contratt}, Dobrushin and Pechersky show that the
specification uniquely determines 
the probability (see Theorem~1 of \cite{do2}). In particular, also the
marginal probability $P_{V}$ on $V$, and the probability
$P_V(\cdot|\dif \vett x)$ conditioned to a vector $\vett 
x$ in $\tilde V$ are determined by $\Gamma$, and the maximum value
taken by $\lambda(\inda,\ind)$ of (\ref{eq:def_lambda}) is bounded
from above.

As a matter of fact, in \cite{do2} the authors do not investigate the
explicit dependence of $\lambda(\inda,\ind)$ on $|\ind-\inda|$, which
is what we need, but such a dependence can be obtained by a slight
modification of their way of reasoning, which leads to the proof of
Theorem~\ref{teor:correlazioni_generico} that appears in 
Appendix~\ref{app:dim_correlazioni}. We need also to introduce 
an upper bound to the mean value of $h$
with respect to the probability $P_V$ on $V$ and to the
probability $P_V(\cdot,\dif \vett x)$, i.e.,
\begin{equation}\label{eq:definizione_h_x}
\langle h\rangle_{\vett x} \equal  \sup_{\ind\in V} \max\left\{ \int_{\mathbb
  R}  h(y) P_\ind (\dif y|\dif \vett x),
\int_{\mathbb R}  h(y) P_\ind (\dif y),C
\right\}\ ,
\end{equation}
where $\vett x\in \tilde V$, while $C$ is the constant entering
Condition~\ref{cond:compattezza}. So we can state
\begin{theorem}[Decay of Correlations]\label{teor:correlazioni_generico}
Let $f$ be a measurable function from $\mathbb R^{|\tilde V|}$ to
$\mathbb R$, depending on the sites lying in the set $\tilde
V$. Let $g$ be a measurable function from $\mathbb R^{|V|}$ to
$\mathbb R$, depending on the sites contained in the set $V$, with
$V\cap \tilde V=\emptyset$, and $h$ be a compact function such that inequality
(\ref{eq:maggiorazione_g_h}) is satisfied.

Let $\Gamma\in \Theta(h,C,\delta)$. Then, there exists a constant $\bar{K}_0$,
depending on $h,C,a,b$ only, such that, if $\Gamma\in
\Delta(h,\bar{K}_0C,\alpha)\cap \Theta(h,C,\delta)$ with any
$\alpha$, one can find constants $D, 
c>0$, for which one has
\begin{equation}\label{eq:correlazioni_generico}
 \left|\langle fg\rangle -\langle f\rangle\langle g\rangle \right|
\le D\left|V\right|^2 \left| \int_{\mathbb R^{|
    V|}}  f(\vett x)\langle h\rangle_{\vett x} P_{\tilde V} (\dif
\vett x)\right| \exp\left(-c d(V,\tilde V)\right)\ .
\end{equation}
The constant $D$ depends on $a,b,\alpha,\delta$ only,  while one has 
\begin{equation}\label{eq:lunghezza_correlazione}
c\equal -\frac{1}{b\range}\log\left[\frac 12\left(\max\{\alpha,\delta
  a^b\}+1\right)   \right] \ .
\end{equation}
\end{theorem}

\subsection{Estimate of the variance of $P_n$}\label{sottosez:stima_sigma_P}
The previous way of proceeding can be fitted to our case by choosing
as specification that given by the Gibbs measure relative to
$H$. As $f$ and $g$, we choose polynomials
in $p$ and $q$, depending on two disjoint sets of sites. In fact, on
account of Lemma~\ref{lemma:coeff_nostro_caso}, we know that the $P_n$
are constituted by a sum of such terms. This way we can study
$\sigma_{P_n}$ and state that it is bounded from above as in the
following
\begin{lemma}\label{lemma:stima_P}
There exist constants $\bar\beta>0$, $\bar\eps>0$, $k'>0$ such that,
for any $\beta>\bar\beta$ and any $\eps<\bar\eps$, one has, for $n<1/\eps$,
\begin{equation}\label{eq:stima_P}
\sigma_{P_n} \le \sqrt N\,n!\left(k'\right)^n \mathcal{D}_n\left(
\sqrt 2\beta\right)^{-1}
\left(\eps+ \beta^{-1} \right)^n \ ,
\end{equation}
where the polynomials $P_n$ are defined by (\ref{eq:definizione_P_n})
and $\mathcal{D}_n$ are given in Lemma~\ref{lemma:coeff_nostro_caso}.
\end{lemma}
\blank \textbf{Proof.} Lemma~\ref{lemma:coeff_nostro_caso}
provides the necessary estimates for the coefficients which appear in
the sum  defining $P_n$ (see
(\ref{eq:scomposizione}--\ref{eq:norma_coeff_P})): we can  write
$
P_n=\sum_{i=1}^N f_i ,
$
where $f_i$ are polynomials depending at most on the sites between
$i-n$ and $i+n$. The variance can be expressed as
$
\sigma^2_{P_n}=\sum_{i,j=1}^N\left(\langle f_if_j\rangle-\langle
f_i\rangle\langle f_j\rangle\right).
$
We then consider the set $\mathcal{S}_1\equal \{(i,j) : |j-i| \le
2n\}$ and $\mathcal{S}_2= \mathcal S_1^c$. The proof
goes on by finding an upper bound separately for the contributions
coming from these two sets: for the latter, we use the methods
developed in the present section, while the terms of the former group are
estimated in a way similar to that of Lemma~\ref{lemma:stima_P_punto}.

We start from $\mathcal S_1$. Firstly, we observe that, in general,
one has
$$
\left|\langle
f_if_j\rangle -\langle f_i\rangle \langle
f_j\rangle\right|\le \sigma_{f_i}\sigma_{f_j}
\le \max \left\{\langle f_i^2\rangle, \langle f_j^2\rangle\right\}\ ;
$$
so that it suffices to evaluate $\sup_i\langle f_i^2 \rangle$ in a way
similar to Lemma~\ref{lemma:stima_P_punto} and to sum over the $j$ with
$|j-i|\le 2n$ to see that the contribution due to the terms in this set is
smaller than
$$
\sum_{i,j\in \mathcal{S}_1}\left|\langle
f_if_j\rangle -\langle f_i\rangle \langle
f_j\rangle\right|\le\mathfrak{C}_1 K^{4n+1} (4n+1) (2n+1)!
(2\omega)^{2n} \beta^{-2}\left( \eps+\frac 1{\beta}\right)^{2n}N
\mathcal{D}^2_n\ .
$$

We come now to $\mathcal S_2$. We will show below that the
specification coming from the Gibbs measure satisfies the hypotheses
of Theorem~\ref{teor:correlazioni_generico}, and we make use of it in
estimating the terms in the set $\mathcal{S}_2$ in the following way. We
separate the terms of a definite degree by writing
$$
f_i=\sum_{l=0}^n\frac{n!}{l!(n-l)!}\eps^{n-l}\sum_{s=1}^{|\mathcal{H}^{n-l,i}_{2l+2}|}
c_{is,l} f_{is}^{(l)}\ ,
$$
in which $f_{is}^{(l)}$ is a monomial in $\mathcal{H}^{n-l,i}_{2l+2}$ and
$\sup_{i,l}\sum_s |c_{is,l}|\le \mathcal D_n$. We
fix an index $i$ and use Theorem~\ref{teor:correlazioni_generico} with
$f=f_i$ and $g=f_{js}^{(l)}$, for every $j\neq i$; then, we sum over
$l$, $s$, $j$ and $i$, subsequently. For each $l$ we choose the compact
function $h_l(x)$ as $|x|^{2l+2}$, which satisfies (\ref{eq:maggiorazione_g_h})
for any $f_{js}^{(l)}$.  We will show that, for $\beta$ large enough
and $\eps$ small enough, one has 
\begin{equation}\label{eq:somma_h}
\sum_{l=0}^{n} \langle
h_l\rangle_{\vett x} \le  k^n n! \frac 1\beta\left(\eps+\frac
1\beta\right)^n\exp\left(\sum_{\indamono=1}^{|\tilde 
  V|}\left(\frac {\beta\eps}{\omega} x_\indamono^2+ 8\eps\sqrt {\frac
{\beta}{2\omega}} \left|x_\indamono\right|
\right)\right)\ ,
\end{equation}
for a suitable constant $k$. Then the sum on $s$ brings in a factor
$\mathcal{D}_n$.  As regards the integration in $\tilde V$, we observe
that we can write 
$$
\beta \frac {1-2\eps}{2} x_\indamono^2- 8\eps\sqrt {\frac
{\beta}{2\omega}} \left|x_\indamono\right|\ge \frac \beta4 x_\indamono^2-1\ ,
$$
provided $\eps$ is small enough. Thus, there exists $\bar k>0$
such that
$$
\left|\langle
f_if_j\rangle -\langle f_i\rangle \langle
f_j\rangle\right|\le \bar k^n (n!)^2 \mathcal D_n^2\frac 1{\beta^2}\left( \eps
+\frac 1\beta\right)^{2n} \exp\left[
  -c(|i-j|-2n-1) \right]\ ,
$$
where the constant $c$ is defined by
(\ref{eq:lunghezza_correlazione}), and is, in the present case, equal
to $\log(4/3)/2$. Since the sum over $j$ of such terms converges as
$N\to \infty$, the proof will be concluded if we show that
(\ref{eq:somma_h}) and the hypotheses of
Theorem~\ref{teor:correlazioni_generico} are satisfied.

We proceed as in the proof of Theorem~2 of paper
\cite{do2}, starting from the explicit form of the conditional
probability distribution given by the Gibbs measure: one has
$$
P_{\indmono,\vett x}(x)= \frac 1{Z_{\vett x}} \exp\left[-\beta\left(
    \frac{\omega x^2}{2}+\frac{x^4}{4\omega^2}+\frac \eps\omega x
\cdot x_{\indmono-1} + \frac  \eps\omega x \cdot x_{\indmono+1}
\right)\right] \ ,
$$
in which there appears the conditional partition function
$$
Z_{\vett x}\equal \int_{\mathbb R} \exp\left[-\beta\left(
    \frac{\omega x^2}{2}+\frac{x^4}{4\omega^2}+\frac \eps\omega x
\cdot x_{\indmono-1} + \frac  \eps\omega x \cdot x_{\indmono+1}
\right)\right] \dif x \ .
$$

As regards Condition~\ref{cond:compattezza}, we consider $\eps<2^{-5}$
fixed and define
$$
\hat
y(\vett x)\equal \max \left\{\left| x_{\indmono-1}\right|,
\left| x_{\indmono+1} \right|\right\}\quad \mbox{and } y(\vett
x)\equal \min\left\{1/ \hat y(\vett x),\sqrt{\beta}\right\}.
$$
So, it is easily proved that, for  $\beta>1$, 
inequality $Z_{\vett x}\ge \bar c y(\vett x)/\beta$, 
holds for some $\bar c>0$ indipendent of $\vett x$, $\beta$ and $\eps$. 
To prove it, it is sufficient to observe
that the integrand of $Z_{\vett x}$ is bounded away from zero if $|x|\le
y(\vett x)/\beta$, and then to integrate over such an interval. We
now show item 2 of Condition~\ref{cond:compattezza} for
$h(x)=|x|^{2l+2}$. We note that
\begin{eqnarray*}
\int_{|x|\ge \hat y(\vett x)/4} h_l(x) P_{\indmono,\vett x}(x)\dif x &\le& \frac
2{Z_{\vett x}} \int_{1/(4 y(\vett x))}^{+\infty} x^{2l+2}
\exp\left(-\beta\frac{ x^2}{4}\right)\dif x\\
&\le& \frac{2\sqrt \beta }{\bar c y(\vett x)}\frac
1{\beta^{l+1}}\int_{\sqrt\beta/(4y(\vett 
  x))}^{+\infty} x^{2l+2} e^{-x^2/4}\dif x 
\ ,
\end{eqnarray*}
which, in turn, is smaller than $(l+1)!(B/\beta)^{l+1} $ for a suitable
constant $B$, independent of $\eps$, $\beta$ and $l$, since the
integral decreases as an exponential function of
$\sqrt{\beta}/y(\vett x)$. Here, we have chosen $(l+1)!$ so
that the previous relation is satisfied independently of $l$.
Furthermore, one has
$$
\int_{|x|\le \hat y(\vett x)/4} h_l(x) P_{\indmono,\vett x}(x)\dif x \le 2^{-4l-4}
\left(\hat y(\vett x)\right)^{2l+2}
$$
and $(h_l(x_{\indmono-1})+h_l(x_{\indmono+1}))/h_l(\hat y(x))\ge 1$, for any
$\vett x$: this implies that 
$$
\int_{\mathbb R} h_l(x)P_{\indmono,\vett x}(\dif x)\le
(l+1)!\left(\frac{B}{\beta}\right) ^{l+1}+
\frac 1{16} h_l\left(x_{\indmono-1}\right)+\frac 1{16} h_l\left(
x_{\indmono+1}\right) \ .
$$
So, item 2 of Condition~\ref{cond:compattezza} holds with
$ C= (l+1)!\left(B/\beta\right) ^{l+1}
$
and $c_1=c_{-1}=1/16$. Since we have $a=2$, $b=2$,
$\range=1$, item 1 of Condition~\ref{cond:compattezza} holds with $\delta
a^b=1/2$.

Condition~\ref{cond:contratt} is proved by computing two limits, first
letting $\beta$ tend to infinity and then letting $\eps\to 0$. In fact,
let $\vett x^{(m)}$, for $m=1,2$, be two different configuration such
that $|x^{(m)}_{\indmono-1}|\le 
\tilde K/e \sqrt{Bl/(e\beta)}$ and $| x^{(m)}_{\indmono+1}|\le\tilde
K/e\sqrt{Bl/(e\beta)}$. Then it is easily checked that
$$
\lim_{\beta\to \infty}\int_{\mathbb R}\left|P_{\indmono,\vett
  x^{(1)}}(x)-P_{\indmono,\vett   x^{(2)}}(x)\right|\dif x =\int_{\mathbb{R}}
\dif z\, e^{-z^2} \left|f(\eps,z,\vett z^{(1)})-f(\eps,z,\vett
  z^{(2)})\right|\ ,
$$
where
$$
f(\eps,z,\vett z^{(m)})\equal\exp\left(-\frac{\eps}{\omega}z\left( z^{(m)}_{i-1}
+z^{(m)}_{i+1}\right) +\frac{\eps^2}{2\omega^2} \left(z^{(m)}_{i-1}+
z^{(m)}_{i+1}\right)^2 \right)
$$
and $z=x\sqrt\beta$, $z^{(m)}_j=x^{(m)}_j\sqrt{\beta}$. Now,
by the dominated convergence theorem, one has that the limit for
$\eps\to 0$ of $f(\eps,z,\vett z^{(m)})$ is equal to 1. Here, use is
made of the fact that $\eps |z^{(m)}_j|$ can be bounded from above by
$\eps \tilde K/e 
\sqrt{Bl/e}\le \tilde K/e \sqrt{B\eps/e}$, because $l\le n$, and $n$
is smaller than $1/\eps$, by hypothesis. So, for $\beta$ sufficiently
large and $\eps$ small enough one has
$$
\int_{\mathbb R}\left|P_{\indmono,\vett x^1}(x)-P_{\indmono,\vett
  x^2}(x)\right|\dif x \le \frac 14 \ .
$$
We have chosen a bound to $\vett x^m$ of this particular form, because 
the constant $\bar{K}$ in Condition~\ref{cond:contratt}  turns out to
be smaller than $\tilde K^{2l+2}$, so that it is independent of $\beta$.
So Condition~\ref{cond:contratt} holds with $\bar K=\tilde K^{2l+2}$,
$k_{-1}=k_1=1/2$ and $\alpha=1/2$. Thus,
Theorem~\ref{teor:correlazioni_generico} holds.

There still remains to estimate $\langle h_l\rangle_{\vett x}$ in our
case. By looking at its definition (\ref{eq:definizione_h_x}), we
notice that we have to estimate the integrals $\int  h(y)
P_\ind (\dif y|\dif \vett x)$ and $\int  h(y) P_\ind (\dif y)$. Now,
on account of Lemma~\ref{lemma:marginale} and  relation
(\ref{eq:rapporto_n_n_tilde}), the distribution functions of $P_\ind
(\dif y|\dif \vett 
x)$ and $P_\ind (\dif y)$ can be bounded by
$$
 K^{2n+2}\mathfrak{C}_2\mathfrak{C}_1\sqrt{\frac{\beta}{2\pi\omega}}
\exp\left(\sum_{\indamono=1}^{|\tilde
  V|} \left(\frac {\beta\eps}{\omega} x_\indamono^2+ 8\eps\sqrt {\frac
{\beta}{2\omega}} \left|x_\indamono\right|\right)\!\right)\!e^{-\beta
  y^2/(2\omega)}\ .
$$
Then, we use the  bound to $C$ previously found, together with the fact that
$$
\sqrt{\frac{\beta}{2\pi\omega}}\int_{\mathbb{R}} y^{2l+2}e^{-\beta
  y^2/(2\omega)}\dif y= \left(\frac{2\omega}{\beta}\right)^{l+1}
\frac{(2l+1)!!} {2^{l+1}}, 
$$
and we get
(\ref{eq:somma_h}). This concludes the proof. 
\begin{flushright}Q.E.D.\end{flushright}\blank

\subsection{Estimate of the variance of
  $X_n$}\label{sottosez:stima_sigma_X} Lemma~\ref{lemma:stima_P} of
Section~\ref{sottosez:stima_sigma_P} enables us to
bound from below the variance of $X_n$ defined by
(\ref{eq:definizione_X_n}) and to estimate the correlation coefficient
$\rho_{X_n,H}$, according to Lemma~\ref{lemma:stima_correlazione},
which we prove here.

\blank \textbf{Proof of Lemma~\ref{lemma:stima_correlazione}}
We start by recalling that, on account of
(\ref{eq:definizione_X_n}), one has $X_n=
-\nucleo_1+ \sum_{j=2}^n P_j$, with $\nucleo_1$ defined by equations
(\ref{eq:determinazione_chi}--\ref{eq:determinazione_Psi}) of
Section~\ref{sez:telchi}. It is easily seen that
$\nucleo_1=F+G+\mathcal{R}_1$, in which
$$
 F\equal -\frac{\eps
}{2 \omega} \sum_{i=1}^{N-1} p_ip_{i+1}\quad \mbox{and }G\equal \frac 3{32
   \omega^2} \sum_{i=1}^N p_i^4 \ ,
$$
and $\mathcal{R}_1$ is the remainder. Then, we study the properties of
$F$ and $G$, for which the mean value, the 
variance and the correlation with $H$ can be computed almost exactly, and we
extend such properties to $\nucleo_1$, and to the whole $X_n$, by
observing that, in some sense, $\nucleo_1$ is the term of first order in
$\eps+\beta^{-1}$.

As regards formula (\ref{eq:minorazione_sigma_X_n}), we notice that,
since $F$ is odd in the momenta, while $G$, $\mathcal{R}_1$ and the
measure are even, then $F$ is uncorrelated both with $G$ and with
$\mathcal{R}_1$. Furthermore, one can observe that
$$
\left\langle
G\,R_1\right\rangle- \left\langle
G\right\rangle\left\langle R_1\right\rangle =
\frac{9}{2^{10} \omega^4} \sum_{i=1}^N \langle q_i^2\rangle \left(\langle
p_i^6\rangle-\langle p_i^4\rangle \langle p_i^2\rangle\right)\ ,
$$
and use the estimates of Lemma~\ref{lemma:marginale} to bound from
above $\langle q_i^2\rangle$, in order to prove that 
$ \sigma^2_{\nucleo_1}\ge \sigma^2_F + \sigma^2_G +2C_{G,\mathcal R_1} \ge
N(\eps^2+\beta^{-2})/(8\beta^2)$,
where the second inequality holds for  $\eps$ and $\beta^{-1}$
small enough. On the other
hand, making use of (\ref{eq:stima_P}) of
Lemma~\ref{lemma:stima_P} together with the estimate for 
$\mathcal{D}_n$ given by (\ref{eq:norma_coeff_P}) of
Lemma~\ref{lemma:coeff_nostro_caso}, one has
$$
\sigma_{X_n} \ge \sigma_{\nucleo_1} - \sum_{j=2}^n \sigma_{P_j} \ge
\sqrt{N}\,\frac {\eps+\beta^{-1}}{4\beta} \left(1- 
 \sum_{j=2}^n (j!)^4 \left(\eps+\beta^{-1}\right)^{j-1}
 \kappa_2^j \right) \ ,
$$
for a suitable constant $\kappa_2$, if $\beta^{-1}$ and $\eps$ are
sufficiently small. Now,  for
$n<\kappa_2^{-1/4}(\eps+\beta^{-1})^{-1/4}$, the sum is
smaller than a constant multiplied by $\eps+\beta^{-1}$ and
this proves (\ref{eq:minorazione_sigma_X_n}).

As for (\ref{eq:stima_correlazione}), we observe that, since $H$ is
even in the momenta, $F$ and $H$ are uncorrelated, so that, using
$\rho_X,Y<1$, one gets
$$
\left|\rho_{X_n,H}\right| \le \frac
1{\sigma_{X_n}\sigma_H}\left(\left|C_{\nucleo_1 -F,H}
\right|+\sum_{j=2}^{n} \left|C_{P_j,H} \right|\right) \le
\frac{\sigma_{\nucleo_1-F}}{\sigma_{\nucleo_1}}\frac{\sigma_{\nucleo_1}}{\sigma_{X_n}} +
\frac{\sum_{j=2}^n\sigma_{P_j}}{\sigma_{X_n}}\ .
$$
As we have just shown, for $n<\kappa_2^{-1/4}(\eps+\beta^{-1})^{-1/4}$
the last term at the r.h.s. tends to zero as $\eps +\beta^{-1}$,
and in the same way behaves $\sigma_{\nucleo_1}/\sigma_{X_n}-1$. So,
we limit ourselves to study $\sigma_{\nucleo_1-F}/\sigma_{\nucleo_1}=
  1/\sqrt{1+\sigma^2_F/\sigma^2_{\nucleo_1-F}}$.
By computing explicitly $\sigma^2_F$ and applying the upper bound
(\ref{eq:stima_P}) to $\sigma^2_{\nucleo_1}\ge
\sigma^2_{\nucleo_1-F}$, we get that there exists a constant  $\bar
\kappa\ge 1$ such that
$$
\frac{\sigma_{\nucleo_1-F}}{\sigma_{\nucleo_1}}\le \left(1+\frac
1{\kappa}\frac{\eps^2}{\left(\eps+ \beta^{-1} \right)^2}
\right)^{-1/2}\ .
$$
Since the r.h.s. differs from 1 by a quantity larger than
$\eps^2\beta^2$, the corrections given by the other terms can
be neglected if $\beta\ge \eps^{-1}$. This
completes the proof.
\begin{flushright}Q.E.D.\end{flushright}\blank

\section{Relation between stability estimates and relaxation
  times}\label{sez:definizione}
In the present section we discuss which implications the existence of an
adiabatic invariant has in the frame of ergodic theory. The main point
is that it can provide a lower bound to the relaxation time to
equilibrium. Since there is no agreement in the literature on the definition of
relaxation time, we will give here a mathematically clear form to such a
concept. To this end, we need a preliminary discussion of an
a priori bound to the time autocorrelations (see
Theorem~\ref{teor:mix}). This is provided in
Section~\ref{sottosez:rilassamento}. Then, in
Section~\ref{sottosez:mescolamento}, we define the concept of
relaxation time.

\subsection{Relaxation times and time
  correlations}\label{sottosez:rilassamento} 
One of the open problems in statistical mechanics is that of
thermalization, i.e., to establish whether a system, starting from a
given microscopic state, does attain thermodynamic equilibrium, and,
if this is the case, to estimate the time scale needed to reach it. Such a
time scale is usually called the \emph{relaxation time}. From a physical point of
view the situation is complicated, because certain systems, for
example gases, reach equilibrium on a very short time scale, while
others, for example glasses, are believed to reach equilibrium on
geological time scales.

Linear response theory (see \cite{green,kubo}) shows that
susceptibilities can be expressed in terms of the time autocorrelations
of suitable dynamical variables (namely, those conjugated to the
perturbing field). In particular, the susceptibilities assume the
equilibrium values only for measurements which last a time large
enough, i.e., larger than the time needed by the time autocorrelations
to become negligible.

Now, the time correlations between pairs of dynamical variables are widely
studied in the case of chaotic systems (see, for example,
\cite{liverani,keller-liv} or the monograph
\cite{chernov}). For such systems,
the correlations are known to tend to zero, as $t\to \infty$, and one of
the problems is to estimate the decay rate, for
long times, of the time autocorrelations of all dynamical
variables. This however amounts to give an upper bound to the time
autocorrelations. From the standpoint of linear response theory, it is
also significant to bound from below the
time autocorrelations of suitably chosen dynamical variables, because
this leads to a lower bound to the relaxation time.

Corollary~\ref{cor:autocorr} of Theorem~\ref{teor:main} gives an
estimate of such a kind, 
showing that, for the system here considered, the relaxation
time is larger than a constant $\bar t$, which is exponentially large
in the perturbation parameters (and moreover does not depend on
the number of degrees of freedom of the system). Results analogous to
Corollary~\ref{cor:autocorr} are of a general type. In fact one has
\begin{theorem}[Bound to the autocorrelation of a dynamical
    variable]\label{teor:mix} 
Suppose that, for a dynamilcal variable $X$, there exists a constant
$\eta > 0$ such that
\begin{equation}\label{eq:ipt}
\left\| [X,H] \right\| \le \eta \sigma_X \ ;
\end{equation}
then one has
\begin{equation}\label{eq:mixst}
   C_X(t) \ge 1- \frac 12  \eta^2 t^2\ .
\end{equation}
\end{theorem}
\blank
\textbf{Remark.} This theorem is a slight modification of
  Theorem~1 of \cite{carati} and is proved in the same
  way. On the other hand, we think that the
 decision to focus on the time autocorrelation, which we make here at
 variance with paper \cite{carati}, is
 crucial, if one aims at obtaining significant estimates in the
 thermodynamic limit.
\blank

\blank \textbf{Proof.}
 Introduce the difference $\delta \equal X_t - X$. As $X_t$
satisfies the Liouville equation and $X$ is time--independent, one
has $\partial_t \delta =$ $ \partial_t X_t = - [H,X_t]$, which in terms
of $\delta$  takes the form
\begin{equation}\label{eq3}
 \partial_t \delta = - [H,\delta] + Y \ ,
\end{equation}
with $Y \equal - [H,X]$. It is well known that, $\mu$ being invariant, the
solutions of the Liouville equation are generated by a one--parameter
group $\hat U(t)$ of unitary operators in the sense that $X_t = \hat
U(t) X $. As $\delta(0)=0$, the solution of equation (\ref{eq3}) is given by
$$
\delta = \int_0^t \hat U(t-s) Y \dif s \ ,
$$
so that, $\hat U$ being unitary, one gets the estimate
$$
\left\|\delta \right\|\le \int_0^t \left\|\hat U(t-s) Y\right\| \dif s = t
\left\|Y\right\|  \le \eta t  \sigma_X\ .
$$
Then, one gets the thesis by using the simple identity
$$
C_X(t) =1 - \frac{\left\| X_t - X\right\|^2}{2\sigma^2_X}\ .
$$
\begin{flushright}Q.E.D.\end{flushright}\blank

\subsection{Definition and evaluation of relaxation
  times}\label{sottosez:mescolamento}
Taking into account the relation between susceptibilities
and time autocorrelations, it is meaningful to introduce a parameter
$a=a(t)$, with values in $[0,1]$, which estimates how much the system
is close to equilibrium, after a finite time $t$. So, for the set
$\mathcal{B}\subset L^2\cap \mathcal{C}^{\infty}$ of the dynamical variables
uncorrelated with $H$, we define
\begin{definizione}[Correlation level]
  The correlation level $a(t)$ at time $t$ is defined as
$
a(t)\equal \sup_{X\in \mathcal{B}}\left|C_X(t) \right|
$.
\end{definizione}
\blank
\textbf{Remark.}
One can limit oneself to the smooth observables, because these are the
physically relevant ones.\footnote{Notice that one can find ``pathological''
functions for which the decay is arbitrarily slow (see \cite{crawford})
even for strongly chaotic systems. For this reason, the control is usually
restricted to a fixed continuity class.}
\blank

In the chaotic case $a$ tends to 0 as $t\to \infty$: thus, looking for
the decay to zero of the correlations is equivalent to looking at the
asymptotic behaviour about zero of $t(a)$, the inverse function 
of $a(t)$.\footnote{As a
  matter of fact, we cannot guarantee that $a(t)$ is invertible, but
  we will give below a meaningful univocal definition of $t(a)$ (see
  definition~\ref{def:2}).} On the other hand, according to linear
response theory, it is more meaningful to look at the time after which
the correlations are below a certain threshold. So, we introduce the
following notion
\begin{definizione}[Relaxation time relative to level $a$]\label{def:2}
The relaxation time relative to level $a$ is defined as
$t(a)\equal \inf t^*(a) ,$
where $t^*(a)$ is such that
$$
\sup_{X\in\mathcal{B}}\left|C_{X}(t) \right|
\le a \quad \mbox{for all } t\ge t^*(a)\ .
$$
\end{definizione}
\blank
\textbf{Remark.} In order to provide a significant lower bound to the
relaxation time 
$t(a)$ at level $a$, as previously defined, it is clearly
sufficient to find the time at which the autocorrelation of at least one
dynamical variable $X$ uncorrelated with the Hamiltonian
is certainly larger than $a$. Now, the following corollary on the
relaxation time descends immediately from Theorem~\ref{teor:mix}:
\begin{corollario}[Bound to the relaxation time]\label{cor:mix}
Suppose there exists a dynamical variable $X\in\mathcal{B}$ and a constant
$\eta > 0$ such that
$\left\| [X,H] \right\| \le \eta \sigma_X$;
then one has
\begin{equation*}
   t(a)\ge \sqrt{2(1-a)} \,\frac 1  \eta\ .
\end{equation*}
\end{corollario}
\blank

The point is that  many Hamiltonian systems of interest for  Solid
State Physics reduce to integrable ones in some limit, while, on the
other hand, for integrable systems one has $t(a)=+\infty$ for any
$a<1$,
since their integrals of motion remain correlated for all
times. The question is then, what is the behaviour of such systems when
the perturbation is small, i.e., to study the
ergodic properties of slightly perturbed (or nearly integrable)
Hamiltonian systems. It is natural to think that there exists a sort
of continuity as the perturbation diminishes. Continuity can in fact
be recovered in terms of the time needed for the system to
reach thermalization (i.e., a sufficiently low correlation
level). This is indeed the case in the system we have considered,
because we can say that, as a consequence of Theorem~\ref{teor:main}, one has
the lower bound
 \begin{equation*} 
   t(a) \ge \sqrt{2(1-a)}\,\frac{\eps}{\kappa} \exp\left(\frac 
   1{\kappa \left(\eps+\beta^{-1}\right)}\right)^{1/4}\ ,
 \end{equation*}
which goes to infinity as both $\eps$ and $\beta^{-1}\to 0$.

\section{Conclusions}\label{sez:conclusione}

In this paper, we have constructed, for the Klein Gordon lattice, an
adiabatic invariant, i.e., a dynamical 
variable whose time derivative is small as a stretched exponential
with the perturbation parameters. Thus our result is similar to those
which are known in 
Hamiltonian perturbation theory in the case of a
finite number of degree of freedom or in the case of an infinite
number of them, but at a fixed total energy (see
\cite{partfinite,enfinita}). The
new feature of the present work is that our theorem remains valid in
the thermodynamic limit, because the given bound turns out
to be independent of the number of particles, and depends 
only on intensive quantities. As a corollary, we bound from below the
stability time of such a model. 



We now add some comments. The first one concerns the fact that in our
model we have two perturbation parameters, $\eps$ and $1/\beta$. We
believe however that the only really relevant parameter is $1/\beta$. Indeed,
at least formally 
the parameter $\eps$ can be arbitrarily decreased by performing a suitable
normal form change of coordinates (\cite{giorg2}), and it seems to us
that such a normal form does not alter in any fundamental feature the
perturbation $H_1$ (i.e., its local character). At the moment, however,
we are unable to say anything definite on this point. 

In any case, while the estimate of Section~\ref{sez:condizionata}
could presumably be applied 
to models more general than that studied in this paper, the
estimates on the marginal probabilities of Section~\ref{sez:marginale}
are especially adapted to our model. It would be important to
improve our method, making it more flexible in order to cover more general
situations. 

In our opinion, the real big problem that remains is the construction of
adiabatic invariants for problems in which small denominators appear.
This problem could be overcome in particular cases (see, for example,
paper~\cite{carati}), but  no precise strategy exists yet for the
general case. We plan to tackle this problem soon.


\subsubsection*{Acknowledgements}
We thank very much Prof. A.~Giorgilli for suggestions regarding
the construction of the adiabatic invariant by making use of
the operator $T_\chi$, and the definition of the norms
of Section~\ref{sez:telchi}.

We also thank Prof. L.~Galgani and D. Bambusi for a careful reading of
the manuscript, useful comments and lucid discussions.

\appendix
\section{Estimates for the construction of the adiabatic
  invariant}\label{app:coeff}
Here we intend to prove Lemma~\ref{lemma:coeff} of
Section~\ref{sez:telchi}. In order to do that, we need to recollect
the usual algebraic properties used in perturbation theory (see
\cite{giorg}), adapted to our norm $\|\cdot\|_+$, defined
by (\ref{eq:norma_coeff}). Such properties are stated in
Lemmas~\ref{lemma:passaggio_complesse}--\ref{lemma:proiezioni} later
on, then the proof of Lemma~\ref{lemma:coeff} is briefly sketched.


In order to develop the perturbation theory, a primary role is played
by the action of the operator $L_0$
 and by the projections on its
kernel and its range (see Section~\ref{sez:telchi}),
and these are more easily discussed in terms of the complex variables
which diagonalize $L_0$. These are implicitly defined by
\begin{equation}\label{eq:coord_complesse}
  q_l=\frac{1}{\sqrt{2}}(\xi_l+i\eta_l)\ , \quad p_l=\frac{1}{
    \sqrt{2} }(\xi_l-i\eta_l)\ ,\quad 1\le l\le N
\end{equation}
and in such variables one has
\begin{equation}\label{eq:autovalori}
L_0\xi^j\eta^k=i\omega(|k|-|j|)\xi^j\eta^k\ .
\end{equation}

We must, however, take into account the fact that the norm
$\|\cdot\|_+$ is not invariant under
such a change of coordinates. In fact, such a norm is formally
well defined also for polynomials depending on the variables $(\xi,\eta)$ 
if, in the definition of
$\mathcal{H}^{r,i}_s$ and $\mathcal{P}_{s,r}$, we simply substitute
for $(p,q)$  the pair $(\xi,\eta)$. In that case, denoting by  $f'$ the
transform of $f$ via (\ref{eq:coord_complesse}), one will have, in
general, $\|f\|_+\neq \|f'\|_+$. On the other hand, the following
lemma,
whose proof is identical to that of  Lemma~A.1 of paper
\cite{giorg},
 enables one to estimate the difference between the norms of the
two functions.
\begin{lemma}\label{lemma:passaggio_complesse}
  Let $f(q,p)$ be in $\mathcal{P}_{s,r}$ and let $f'(\xi,\eta)$ be the
  transform of $f$ via (\ref{eq:coord_complesse}). Then, one has $f'\in
  \mathcal{P}_{s,r}$ and
$  \left\|f'\right\|_+\le 2^\frac s2  \left\|f\right\|_+.$
  Moreover, let $g'(\xi,\eta)$ be in $ \mathcal{P}_{s,r}$ and let
  $g(q,p)$ be the transform of $g$ via the inverse of
  (\ref{eq:coord_complesse}). Then, one has $g\in
  \mathcal{P}_{s,r}$ and
$
  \left\|g\right\|_+\le 2^\frac s2  \left\|g'\right\|_+.
$
\end{lemma}

We need also the following lemmas
\begin{lemma}\label{lemma:par_Poisson}
Let $f$ be in $\mathcal{P}_{s,r}$ and $g$ in
$\mathcal{P}_{s',r'}$. Then, $[f,g]\in\mathcal{P}_{s+s'-2,r+r'}$ and
one has, both in real and in complex variables, the inequality
$$
\left\|[f,g]\right\|_+\le (2r+2r'+1)s s'\left\|f\right\|_+ \left\|g
\right\|_+\ .
$$
\end{lemma}
\blank \textbf{Proof.} See Lemma~A.2 of \cite{giorg},
noticing that, for any
fixed $i$, each term of $f$ contained in $\mathcal{H}^{r,i}_s$ has
Poisson bracket different from 0 only with the monomials
of $\mathcal{H}^{r',k}_{s'}$ such that $|i-k|\le r+r'$.
The number of such monomials appearing in the decomposition of $g$
is smaller than $2r+2r'+1$.
\begin{flushright}Q.E.D.\end{flushright}\blank

\begin{lemma}\label{lemma:proiezioni}
  Let $f\in\mathcal{P}_{s,r}$ be a polynomial in complex
  variables. Then $\Pi_\mathcal{N}f$, $\Pi_\mathcal{R}f$ and 
  $L_0^{-1}\Pi_\mathcal{R}f$ belong to $\mathcal{P}_{s,r}$ and the
  following inequalities hold:
$$
  \left\|\Pi_\mathcal{N}f\right\|_+ \le
  \left\|f\right\|_+\ , \quad
  \left\|\Pi_\mathcal{R}f\right\|_+ \le \left\|f\right\|_+
  \ ,\quad
  \left\|L_0^{-1}\Pi_\mathcal{R} f\right\|_+ \le  \left\|
  f\right\|_+  \ .
$$
\end{lemma}
\blank \textbf{Proof.} The fact that  $L_0^{-1}f$ belongs to
$\mathcal{P}_{s,r}$ comes directly from Lemma~\ref{lemma:par_Poisson},
as $H_0$ is in $\mathcal{P}_{2,0}$. The remaining statements are a
consequence of the fact that $L_0$ is diagonal in complex coordinates
and that the smallest eigenvalue of $L_0$ on $\mathcal{R}$ has modulus
$\omega \ge 1$, in virtue of (\ref{eq:autovalori}).
\begin{flushright}Q.E.D.\end{flushright}\blank

\blank \textbf{Proof of Lemma~\ref{lemma:coeff}} We pass to complex variables
  via Lemma~\ref{lemma:passaggio_complesse}
  and proceed by induction on $n$,  checking at each step 
even two supplementary inductive hypotheses:
\begin{itemize}
\item[i)] $\Psi_n$ can be decomposed as $
\Psi_n=\sum_{l=0}^n \Psi_n^{(l)} ,$
where $\Psi_n^{(l)}\in \mathcal{P}_{2l+2,n-l}$;
\item[ii)] the following bound holds 
$$\left\|\Psi_n^{(l)}\right\|_+\le 2^n2^{10(n-1)}
\left(n!\right)^2(n-1)!\frac{n!}{l!(n-l)!}\eps^{n-l}\ .$$
\end{itemize}
As a matter of facts, on account of Lemma~\ref{lemma:proiezioni}, such
an estimate enables one to control the contributions due to $\chi_n$
and $\nucleo_n$, which appear in the recurrent procedure that determines
$\chi_s$, for $s\ge n$. Then, we come back to real variables
via lemma~\ref{lemma:passaggio_complesse} again.
\begin{flushright}Q.E.D.\end{flushright}\blank

\section{Technical proofs}
\subsection{Proof of Lemma~\ref{lemma:periodico}}\label{app:dim_lemma_periodico}
We start by  proving formula
(\ref{eq:maggiorazione_frazione_Q_N}). On account of the symmetry of
the periodic system, one can pass from a  system with $N-1$ particles to
one with $N$ by inserting one more particle after the $i$--th
site, for $i=1,\ldots,N-1$. The potential energy of the corresponding
system is given by
$$
U_N(q_1,\ldots,q_N,q)= U_{N-1}-\frac
\eps{2\omega}\left(q_{i+1}-q_i\right)^2 + \frac \eps{2\omega}
\left(q-q_i\right)^2+ \frac \eps{2\omega}\left(q- q_{i+1}\right)^2+
\frac {q^2}{2\omega}+ \frac{q^4}{4\omega^2}\ .  
$$
Neglecting the second term at the r.h.s. (which gives a contribution
to the partition function which can be bounded from below by 1, and
averaging over $i$ in order to get a traslational invariant system,
one gets
\begin{equation}\label{minorazione_1}
\begin{split}
\frac{Q_N}{Q_{N-1}} \ge& \frac{1}{N-1}\sum_{i=1}^{N-1}
\int_{-\infty}^{+\infty}\dif q_1\ldots
\int_{-\infty}^{+\infty} \dif q_{N-1}\,\tilde{D}_{N-1}(q_1,\ldots,q_{N-1})
\times \\ 
&\times \int_{-\infty}^{+\infty} \!\!\!\dif q
\exp\!\left[\!-\frac{\beta}{2\omega} \left(\!
  q^2+\frac{q^4}{2\omega}+ \eps(q-q_i)^2+\eps(q-q_{i+1})^2\right)\!\right]\ .
\end{split}
\end{equation}
Here we have put $q_N=q_1$. Then, we introduce the function
$\varphi_{q_i}(q)\equal
1-\exp\left[-\beta\eps(q-q_i)^2/(2\omega)\right], $
for which the inequality
$$
\exp\left[-\frac{\beta\eps}{2\omega}(q-q_i)^2+\frac{\beta\eps}{2\omega}
  (q-q_{i+1})^2\right]\ge 1
-\varphi_{q_i}(q) -\varphi_{q_{i+1}}(q)
$$
holds. We will show now that $\varphi_{q_i}(q)$ is small except for a
set of small measure. 
Making use of the previuos inequality, relation (\ref{minorazione_1})
becomes 
\begin{equation}\label{minorazione_2}
\begin{split}
\frac{Q_N}{Q_{N-1}} \ge a(\beta,\eps)- \int_{-\infty}^{+\infty}\dif
q_1\ldots \int_{-\infty}^{+\infty} \dif
q_{N-1}\,\tilde D_{N-1}(q_1,\ldots,q_{N-1})\times\\ \times
\int_{-\infty}^{+\infty} \dif q\, \frac{2}{N-1}\sum_{i=1}^{N-1}
\varphi_{q_i}(q)\exp\left[-\frac{\beta}{2\omega}\left(q^2+\frac{q^4}{2
    \omega}\right)\right]
\ ,
\end{split}
\end{equation}
in which the function $a(\beta,\eps)$ is defined by\footnote{Remark
  that the function $a(\beta,\eps)$ depends on $\eps$ only via
  the term $\omega=\sqrt{1+2\eps}$.} 
$$
a(\beta,\eps)\equal \int_{-\infty}^{+\infty}\dif q\exp\left[-\frac{\beta}{2\omega}\left(
  q^2+\frac{q^4}{2\omega} \right)\right] =\frac{\sqrt{2\omega}
 \, e^\frac{\beta}{8}} {2} 
K_{\frac{1}{4}} \left(\frac{ \beta}{8} \right)\ ,
$$
where $K_\alpha (x)$ is the Bessel modified function of second
kind. The well known properties of $K_\alpha (x)$ imply that
$a(\beta,\eps)$ can be written as 
$a(\beta,\eps) = G(\beta,\eps)\sqrt{2\pi\omega/\beta},$
where $G$ is a function always smaller than 1, approaching 1, at fixed
$\eps$, as $\beta\to+\infty$.
%
%
We go on by dealing with the integral in (\ref{minorazione_2}), first
giving an upper bound for the innermost integral over $q$. We
estimate it by splitting the phase space of the $N-1$ particles
periodic system in two sets: we will fix $\kappa>0$ and consider
$\Omega(N-1,\kappa)$, which is defined  by
\begin{equation}\label{eq:definizione_Omega}
\Omega(N-1,\kappa)\equal \left\{(q_1,\ldots,q_{N-1}) \mbox{ such that }
\sum_{i=1}^{N-1}q_i^2< \frac{2\omega}{\beta}\kappa(N-1)\right\} \ ,
\end{equation}
and its complement. In the latter set, the integral is simply bounded
from above by $2a(\beta,\eps)$.
 On the other hand, in order to estimate the
integral in the set $\Omega(N-1,\kappa)$, we observe that, for any $\kappa_1$,
the number of particles for which
$\left|q_i\right|\ge \sqrt{\kappa_1\kappa2\omega/\beta}$ holds cannot exceed
$(N-1)/\kappa_1$. For these particles the integral is estimated again
by $2a(\beta,\eps)$. For the purpose of estimating the contribution of
the remaining particles, we introduce the function
$$
I(\beta,\eps,\kappa,\kappa_1) \equal \frac{1}{a(\beta,\eps)}
\sup_{\left|y\right|<\sqrt{\kappa_1\kappa2\omega/\beta}} 
\int_{-\infty}^{+\infty}
\varphi_y(q)\exp\left(-\frac{\beta}{2\omega}q^2\right)\, \dif 
q\ .
$$
We point out that
$I(\beta,\eps,\kappa,\kappa_1)$ tends to 0 as 
$\eps$ tends to 0, for $\beta,\kappa,\kappa_1$ fixed.
Then, in the  region
$\Omega(N-1,\kappa)$, for any $\kappa_1> 1$, one has the bound
\begin{equation*}
\int_{-\infty}^{+\infty} \!\!\dif q\, \frac{2}{N-1}\!\sum_{i=1}^{N-1}
\varphi_{q_i}(q)\exp\left[-\frac{\beta}{2\omega}\!\left(q^2+\frac{q^4}{2
    \omega}\right)\right] \!
\le\!
 \left [\frac 2{\kappa_1} + 2 I(\beta,\eps,\kappa,\kappa_1)\right]\!
a(\beta,\eps) \ .
\end{equation*}
We notice that we have provided estimates independent of $q_i$,  so
the integrals over $q_1,\ldots,q_{N-1}$ appearing in
(\ref{minorazione_2}) can simply be estimated as the product of these
upper bounds times the measures of the sets in which the bounds hold.
Now,
we observe that the measure of
$\Omega^c(N-1,\kappa)$
is estimated by
$$
\int_{\Omega^c(N-1,\kappa)}\dif q_1\ldots  \dif q_{N-1}\,\tilde
D_{N-1}(q_1,\ldots,q_{N-1}) \le 
\frac {R_{N-1}(\beta,\kappa)}{Q_{N-1}}
$$
where the function $R_{N-1}(\beta,\kappa)$ is defined by 
\begin{equation}\label{eq:definizione_resto}
\begin{split}
R_{N-1}(\beta,\kappa) &\equal  \int_{\Omega^c(N-1,\kappa)} \dif q_1\ldots
  \dif q_{N-1}\, \exp \left( 
-\frac{\beta}{2\omega}\sum_{i=1}^{N-1} q_i^2\right)\\
&= \left(\frac{2\pi\omega}{\beta} \right)^\frac{N-1}{2} \Gamma\left(
\frac{N-1}{2},\kappa (N-1)\right)
\end{split}
\end{equation}
and $\Gamma(s,x)$ is defined by (\ref{eq:gamma_reg}). This way one
obtains, finally, 
\begin{equation}\label{minorazione_frazione_finale}
\frac{Q_N}{Q_{N-1}}\ge \left(1-\frac 2{\kappa_1}-2I(\beta,\eps,\kappa,\kappa_1)-
2\frac{R_{N-1}(\beta,\kappa)}{Q_{N-1}}\right) a(\beta,\eps)
\ .
\end{equation}

From this expression one can
prove (\ref{eq:maggiorazione_frazione_Q_N}) by induction on $N$. 

\vskip 0.5em
\noindent
We now come to the proof of (\ref{eq:minorazione_P_N}).
We make use of the trivial inequality
$
\mbox{\bf{P}}(A_1\cap \ldots \cap A_r)\ge 1-\sum_{i=1}^r
\mbox{\bf{P}}(A_i^c)
$,
which holds for any probability and any collection of sets
$A_1,\ldots,A_r$. Consequently,  we obtain
$$
\mbox{\bf{P}}_N\left(|q_1|\!<\Theta\sqrt{\frac{2\omega}\beta}
\;\wedge\;\ldots\;\wedge\; |q_r|\!<\! \Theta\sqrt{\frac{2\omega}
  \beta} \right) \ge 1-r\cdot \mbox{\bf{P}}_N \left(|q_1| \!\ge\!
\Theta\sqrt{\frac{2\omega} \beta}\right)\ . 
$$
because, due to the translation invariance of the periodic system,
every set has the same measure. Recall that ${\bf{P}}_N \Big (|q_1| \!\ge\!
\Theta\sqrt{2\omega/\beta}\Big)$ is just the integral of
$\tilde D_N$ times ${\bf{1}}_{|q_i|\ge\Theta \sqrt{2\omega/\beta}}$.
A bound to this integral can be found proceeding as above, i.e., by
symmetrizing on $q_i$,  fixing $\kappa>0$ and integrating
separately over $\Omega(N,\kappa)$ and  its complement
(recall that $\Omega(N,\kappa)$ is defined by
(\ref{eq:definizione_Omega})). This way we get
\begin{eqnarray*}
\mbox{\bf{P}}_N\left(|q_1| \ge \Theta\sqrt{\frac{2\omega}
  \beta}\right) &\le& \frac 1{NQ_N}\sum_{i=1}^N \int_{\Omega(N,\kappa)}
     {\bf{1}}_{|q_i|\ge\Theta \sqrt{2\omega/\beta}} \tilde
     D_N(q_1,\ldots,q_N)\\
& &
      + \frac 1{Q_N}R_N(\beta,\kappa)\ ,
\end{eqnarray*}
where $R_N$ is defined by (\ref{eq:definizione_resto}), and we bound
${\bf{1}}_{|q_i|\ge\Theta \sqrt{2\omega/\beta}}$ by 1 in
$\Omega^c(N,\kappa)$. It is straightforward to notice that the number
of sites for which 
$|q_i|\ge \Theta\sqrt{2\omega/\beta}$, in the interior of
$\Omega(N,\kappa)$, cannot exceed $N \kappa/\Theta^2$. Therefore, the
former term at the r.h.s. of the previous formula is smaller than $1/4r$ if
$\Theta\ge 2\sqrt{\kappa r}$. As far as the latter is concerned, 
we can choose $\kappa$ such that
$R_N(\beta,\kappa)/Q_N\le 1/4r,
$
as we have shown above. For example, we can fix $\kappa = \log
(4rK_0)$. This suffices to infer that, for $\Theta\ge 2\sqrt{r \log (4r K_0)}$, 
(\ref{eq:minorazione_P_N}) is valid.

\subsection{Proof of Lemma~\ref{lemma:rapporto_Z_Q}}\label{app:dim_lemma_rapporto}
The first inequality in (\ref{eq:minorazione_rapporto_Z_Q}) comes
directly from the fact that the integrand appearing in
the definition of $Q_M$ is smaller than the
function $n_{M,\xgot}$, i.e., the
integrand in the definition of  $\bar{\mathcal{Q}}^\xgot_M$.

As regards the second inequality in
(\ref{eq:minorazione_rapporto_Z_Q}), we note that the integrand of
$Q_M$ is equal to the one of $\bar Z^\xgot_M$ multiplied by $\xgot$ terms
of the form $\exp(-\beta\eps q_{m_i} q_{m_{i+1}}/\omega)$ at the
sites, in number $2\xgot$, on the boundary of the blocks, which we denote by
$m_1,\ldots,m_{2\xgot}$, with the convention that $m_{2\xgot+1}=m_1$. Then, we
integrate only in the region in which
the $q$ coordinate of each of these sites is smaller than
$\Theta\sqrt{2\omega/\beta}$, with
$\Theta=2\sqrt{2\xgot\log (8\xgot K_0)}$, and we observe that
\begin{eqnarray*}
\bar Z^\xgot_M &\ge& \exp\left(-4\xgot\eps \Theta^2\right) Q_M
\mbox{\bf{P}}_M\left(|q_{m_1}|\!<\!\Theta\sqrt{2\omega/\beta}
\wedge\ldots\wedge |q_{m_{2\xgot}}|\!<\!
\Theta\sqrt{2\omega/\beta} \right) \\
&\ge& \frac{Q_M}{2}\left(8\xgot K_0\right)^{-32 \eps_0 \xgot^2}\ .
\end{eqnarray*}
Here, $Q_M$ comes from the normalization of the probability, and in
the second line use is made of Lemma~\ref{lemma:periodico}.

We come now to inequalities (\ref{eq:maggiorazione_rapporto_Z_Q}) and
observe at once that the first one is trivial, because, on account of
the identity in (\ref{eq:rapporto_n_n_tilde}), one has $\tilde n_{M,\xgot}\le
n_{M,\xgot}$. The second one is more 
complicated: we begin by proving it in the case in which each block is
constituted by an even number of elements. 

In order to estimate $\mathcal{Q}^\xgot_M$, we divide again the phase space of
the system in the region  $\tilde\Omega$ in which
$|q_{m_1}|<\sqrt{2\omega\kappa/
\beta},\ldots,|q_{m_{2\xgot}}| < \sqrt{2\omega\kappa/
\beta}$, where $\kappa>0$ is a constant to be determined, and in its
complement $\tilde\Omega^c$. The integral over $\tilde\Omega$ is smaller than
$ Q_M\cdot\exp\left(4\xgot\eps\kappa \right),$
while, as regards the complement, we notice that it is contained in the
set in which $\sum_{i=1}^{2\xgot} q_i^2\ge
2\kappa\omega/\beta$. Thus, the integral over such a region is bounded
from above by
$
\mathcal{Q}^{\xgot_1}_{M-2\xgot}\left(2\pi\omega/\beta\right)^\xgot
\Gamma(\xgot,\kappa)$, with $\xgot_1\le\xgot,$
where we have dropped some positive term in the potentials, then we
have integrated first over $q_{m_1},\ldots,q_{m_{2\xgot}}$ (which gives
the term $\left( 2\pi\omega/\beta\right)^\xgot
\Gamma(\xgot,\kappa)$, with $\Gamma(\xgot,\kappa)$ defined by
(\ref{eq:gamma_reg})); then the blocks made of just 2 particles
disappear, so that the integration over the remaining positions gives
the term $ \mathcal{Q}^{\xgot_1}_{M-2\xgot}$. This way we get
$$
\mathcal{Q}_M^\xgot\le Q_M\cdot\exp\left(4\xgot\eps
\kappa\right) +\left( \frac {2\pi\omega}\beta\right)^\xgot \Gamma(\xgot,
  \kappa) \cdot\mathcal{Q}_{M-2\xgot}^{\xgot_1}\ ,\quad\mbox{with}\quad
\xgot_1\le\xgot\ . 
$$

Now, we apply the previous inequality to the function
$\mathcal{Q}_{M-2\xgot}^{\xgot_1}$ at the r.h.s, and we end up with a
relation similar to the previous one, in which however there appears
the function $\mathcal{Q}_{M-2\xgot-2\xgot_1}^{\xgot_2}$, with $\xgot_2\le
\xgot_1$. So, we can iterate this procedure, observing that
$\Gamma(\xgot,\kappa)$ is an increasing function of $\xgot$, and we get
$$
\mathcal{Q}_M^\xgot\le \exp\left(4\xgot\eps\kappa\right)
\sum_{j=0}^J\left( \frac {2\pi\omega}\beta\right)^{\sigma_j} Q_{M-2\sigma_j}
\left(\Gamma(\xgot,\kappa)\right)^j\ , \quad\mbox{with}
\quad Q_0\equal 1\  ,
$$
where we define $\sigma_j=\sum_{k=0}^j
\xgot_k$, with $\xgot_0=\xgot$, and $J$ represents the integer such that
$\sigma_J=M/2$. We make use of inequality
(\ref{eq:maggiorazione_frazione_Q_N}) of Lemma~\ref{lemma:periodico}
and finally get, if the series converges,
$
\mathcal{Q}_M^\xgot\le \exp\left(4\xgot\eps\kappa\right)
Q_M \sum_{j=0}^\infty (K_0^{2\xgot} \Gamma(\xgot,\kappa) )^j 
$.
We point out that the common ratio of this geometric series is a decreasing
function of $\kappa$, which tends to 0 as $\kappa\to+\infty$: thus, we
choose $\bar \kappa=\bar\kappa(\xgot,K_0)$ so as to satisfy
(\ref{eq:determinazione_kappa_barra}), and obtain the relation
\begin{equation}\label{eq:caso_pari}
\mathcal{Q}_M^\xgot\le 2\exp\left(4\xgot\eps\bar\kappa(\xgot,K_0)\right)
Q_M\ .
\end{equation}

If a number $\lambda\le\xgot$ of blocks is constituted by
an odd number of elements, we integrate on one of the sites on the
boundary of each of these blocks, in order that each of the blocks, in
number $\xgot'$, of the resulting lattice contains an even number of
elements. By dropping some suitably chosen interaction terms in the
potential, one gets
$
\mathcal{Q}^\xgot_M\le
\left(2\pi\omega/\beta\right)^{\lambda/2}\mathcal{Q}^{\xgot'}_{M-\lambda},$
with  $\xgot' \le \xgot$,
where the blocks made of just one particles disappear. Now
we can use (\ref{eq:caso_pari}) with $Q_{M-\lambda}$ instead
of $Q_M$. Then, making use of Lemma~\ref{lemma:periodico} to
express $Q_{M-\lambda}$ in terms of $Q_M$, we get
(\ref{eq:maggiorazione_rapporto_Z_Q}).

\subsection{Proof of Theorem~\ref{teor:correlazioni_generico}}
\label{app:dim_correlazioni}
As already said, the proof is performed by bounding from above  every
term at the r.h.s. of (\ref{eq:diff_misure_2}), i.e.,
$\lambda(\inda,\ind)$, for $\inda,\ind\in V$.

We point out that the expectations in the definition of
$\lambda(\inda,\ind)$ depend on the choice of $Q$, which
is not completely fixed by its marginal probabilities (see comments on
relations (\ref{eq:congiunta})).
In fact, the main part of paper \cite{do2} consists in introducing a
suitable reconstruction operator (on the space of the joint probabilities) 
which enables one to find a joint probability distribution that
minimizes $\lambda$, starting from an initially chosen one. We adopt
the same technique, with the only difference that we
apply it not at all sites of the lattice, but only at those lying on
the complement of a fixed set $\bar V$ (we will call it $\bar
\T\equal \T\backslash\bar V$).

We also need to control, together with $\lambda(\inda,\ind)$, the
auxiliary quantity 
$$
\gamma(\ind) \equal \mbox{\bf E} \left[
{\bf 1}_{\xi_\ind^1\neq \xi_\ind^2} \right]\ ,
$$
where $\xi^1$ and $\xi^2$ are the same Gibbsian
fields entering in (\ref{eq:def_lambda}).

We introduce, then, the main tool of the proof, i.e., the
reconstruction operator
$U_\ind$, with $\ind\in \T$, which will enable us to construct the
joint probabilities $Q(\dif \vett x,\dif \vett y)$ on $V$ (see
formula (\ref{eq:diff_misure_2})). This operator is defined  
on a couple of fields $(\xi^1,\xi^2)$ having the same
conditional probability at $\ind$, as follows. For each pair of configurations
$\vett x^1,\vett x^2\in \mathbb R^{|T|}$, we denote by
$P^\ind_{\vett x^1,\vett x^2}$ the measure on $\mathbb R^2$ for which
the minimum of the distance between $P_{\ind,\vett x^1}$ and $P_{\ind,\vett
x^2}$ is attained, i.e., such that,
for any measurable $B\subset \mathbb R$, one has
\begin{equation*}
\begin{split}
&P^\ind_{\vett x^1,\vett x^2}(\mathbb R\times B) = P_{\ind,\vett
  x^1}(B)\ ,\quad 
P^\ind_{\vett x^1,\vett x^2}(B\times \mathbb R) = P_{\ind,\vett
  x^2}(B)\ ,\\
&\mbox{and }\int_{\mathbb R^2}{\bf 1}_{x\neq y} P^\ind_{\vett
  x^1,\vett x^2}(\dif x,\dif y)= 
D\left(P_{\ind,\vett x^1},P_{\ind,\vett x^2}\right)\ .
\end{split}
\end{equation*}
Such a definition enables us to describe the action of $U_\ind$,
because this operator maps the couple $(\xi^1,\xi^2)$ into
$(\hat\xi^1,\hat\xi^2)$ such that, for any measurable $C\subset \mathbb R^2$,
$$
P\left(\left(\hat\xi^1_\ind,\hat\xi^2_\ind \right)\in C \,|\,
\hat\xi^1_{\T\backslash \{\ind\}}=\vett x^1_{\T\backslash\{\ind\}}, \hat
\xi^2_{\T\backslash \{\ind\}}=\vett x^2_{\T\backslash
  \{\ind\}}\right)=P^\ind_{\vett x^1,\vett x^2}(C)\ ,
$$
and, for any finite $V\subset \T$ not containing $\ind$,
the joint probability distribution of $(\hat\xi^1_V,\hat\xi^2_V)$
coincides with that of $(\xi^1_V,\xi^2_V)$.

The effect of $U_\ind$ on $\gamma(\ind)$ and $\lambda(u,\ind)$ is
described in detail in Lemmas~2, 3 and 4 of the work
\cite{do2}. Following such a paper, we adopt the convention that the
quantities relative to
the reconstructed couple $(\hat\xi^1,\hat\xi^2)$ are distinguished
from the corresponding ones relative to $(\xi^1,\xi^2)$ by adding the
symbol \^{}. For every set $S\subset T$ we define the operator
\begin{equation}\label{eq:definizione_U_S}
U_S\equal U_{\ind_1}\circ U_{\ind_2}\circ\ldots\circ U_{\ind_m}\ ,
\end{equation}
where the order of the points $\ind_1,\ldots,\ind_m$, contained in
 $S\cup \partial_{b\range} S$ is
 chosen in a suitable way. This is described in full detail in the
 proof of the following Lemma~\ref{lemma:ricostruzione}.
 If we define 
$
\gamma_S\equal \sup_{\ind\in S} \gamma(\ind)$ and $
\lambda_S\equal  \sup_{\inda,\ind\in S}\lambda(\inda,\ind),
$
we can describe the action of $U_S$ on a couple of fields having the
same conditional probability on $S\cup \partial_{b\range} S$ accordingly
to the following lemma, which is proved in Appendix~\ref{app:condiz}.
\begin{lemma}\label{lemma:ricostruzione}
Let $(\xi^1,\xi^2)$ be a couple of fields having the
same conditional probability on $S\cup \partial_{b\range} S$, given by
a specification $\Gamma\in \Theta(h,C,\delta)\cap
\Delta(h,\bar{K}C,\alpha)$, and 
$(\hat\xi^1,\hat\xi^2)=U_S(\xi^1,\xi^2)$. Then, one has
$$
\left(\begin{array}{c}
\hat\gamma_S\\ 
\hat \lambda_S
\end{array}\right) \le 
A\left(\begin{array}{c}
\gamma_{S\cup \partial_{b\range} S}\\ 
\lambda_{S\cup \partial_{b\range} S}
\end{array}\right)\ ,
$$ 
in which the matrix $A$ is defined by
$$
A\equal \left(\begin{array}{cc}
\alpha +N\bar{K}^{-1}& C^{-1}M\bar{K}^{-1}\\
C\left(R+N\bar{K}^{-1}\right) & \delta a^b+M\bar{K}^{-1}
\end{array}
\right)\ ,
$$
where $N,M$ and $R$ are constants depending on $a$ and $b$ only.
\end{lemma}

We remark that, if the eigenvalues of $A$ are smaller than 1, the
reconstructed quantities are smaller than the initial ones. So, we
want to iterate the reconstruction procedure as much as possible. It
turns out that we can iterate the procedure at most a number of times
proportional to the distance between $V$ and $\tilde V$. The reason
is the following.

In our case, $\xi^1$ is the field relative to the equilibrium
Gibbs measure and $\xi^2$ that relative to the probability
conditioned to the configuration $\vett x$ on $\tilde V$, which we
consider as fixed. It is apparent that such fields have the same
conditional probability on every set which does not intersect $\tilde
V$, but not on the whole $\T$; by hypothesis, this conditional probability is that
given by $\Gamma\in \Theta(h,C,\delta)\cap
\Delta(h,\bar{K}C,\alpha)$. Since the reconstuction procedure shrinks
the set $S$ on which we can control $\gamma$ and $\lambda$, we can iterate
it until $V\subset S$. So, the maximum number of iterations is
attained if we start by reconstructing on $V\cup \partial_{n b \range}
V$, where $n$ is the largest number such that $ \partial_{(n+1) b \range}
V\cap \tilde V=\emptyset$. We use Lemma~\ref{lemma:ricostruzione} as
the first step of a recurrent scheme, by applying each time
$U_{V_m}$, where $V_{m+1}=V_m\cup\partial_{b\range}
\tilde V_m$,  $V_0=V$. In virtue of Lemma~\ref{lemma:ricostruzione},
after the application of $U_{V_m}$, one has
$$
\left(\begin{array}{c}
\hat\gamma_{V_m}\\ 
\hat \lambda_{V_m}
\end{array}\right) \le 
A\left(\begin{array}{c}
\gamma_{V_{m+1}}\\ 
\lambda_{V_{m+1}}
\end{array}\right)\ .
$$
Thus we get that the
final values of $\gamma_V$ and $\lambda_V$ are smaller than the result
of the application of the matrix
$A^n$ to the vector with components $\gamma$,
$\lambda$. Moreover, we observe that we can write $A=J^{-1}\tilde{A}
J$, where $\tilde A$ and $J$ are defined by
$$
\tilde A\equal \left(\begin{array}{cc}
\alpha +N\bar{K}^{-1}& M\bar{K}^{-1}\\
R+N\bar{K}^{-1} & \delta a^b+M\bar{K}^{-1}
\end{array}
\right) \quad\mbox{and }
J\equal \left(\begin{array}{cc}
C& 0\\
0 & 1
\end{array}
\right)\ .
$$
This way we get $A^n=J^{-1}\tilde A^nJ$. As the component
$\lambda_V$, which is the one we are interested in, is not affected by
the action of $J^{-1}$, we can write that it is the second component
of the matrix product
$$
\left.\tilde A\right.^n \left(\begin{array}{c}
  C\gamma\\ \lambda\end{array}\right)\ . 
$$
Since the eigenvalues of $\tilde A$ are smaller than
$
G\equal \max\{(\alpha+1)/2,(\delta a^b+1)/2\}<1,
$
if $\bar{K}$ is large enough, there exists $\bar{K}_0$ such that
$$
\lambda_V\le (D/2)\max\{\lambda,C \gamma\}
G^n\ ,
$$
where $D$ is a constant depending on $a,b,\alpha,\delta$ only.
On the other hand, $n=d(V,\tilde V)/(b\range)$ and $\gamma\le
1$, from which 
there follows
$$
\lambda_V\le \frac D2\langle h\rangle_{\vett
  x} \exp\left(-c\,d(V,\tilde V)\right)\ ,
$$
where $c$ is defined in (\ref{eq:lunghezza_correlazione}).

In order to show (\ref{eq:correlazioni_generico}), we need only
the use of (\ref{eq:diff_misure_2}) in estimating the term in
brackets of (\ref{eq:correlazioni_condizionato}). We then observe that
the r.h.s. of (\ref{eq:diff_misure_2}) is smaller than $2 |V|^2
\lambda_V$, for the joint probability we have just found, and this
concludes the proof.

\subsection{Proof of Lemma~\ref{lemma:ricostruzione}}\label{app:condiz}
Lemma~5 of work \cite{do2} shows the result of the application of $U_\ind$, in a
suitably chosen order, to every site of $\T$ in sequence: one obtains
that, for the couple of fields $(\xi^1,\xi^2)$, with the same
specification $\Gamma\in \Theta(h,C,\delta)\cap
\Delta(h,\bar{K}C,\alpha)$ and $\bar{K}\ge 1$, and for the reconstructed couple
$(\hat\xi^1,\hat\xi^2)$ the
following matrix relation holds\footnote{As a matter of fact, $A$
  is not the same matrix 
which appears in \cite{do2}, since we needed to make the dependence on
$C$ explicit. Our statement can be proved by checking, in proving the
induction (11)-(14) of \cite{do2}, that the constants $N(\cdot,\cdot)$
are proportional to $C^2$, the constants $N(\cdot)$ and
$M(\cdot,\cdot)$ are proportional to $C$ and the constants $M(\cdot)$
are independent of $C$.}
\begin{equation}\label{eq:matrice_completo}
\left(\begin{array}{c}
\hat\gamma_T\\ 
\hat \lambda_T
\end{array}\right) \le 
A
\left(\begin{array}{c}
\gamma_T\\ 
\lambda_T
\end{array}\right)\ .
\end{equation}

In the proof of Lemma~5 of \cite{do2}, the order of the $U_\ind$'s
is chosen in the following way: the lattice is partitioned in $b$ disjoint
sublattice, $Z_0,\ldots, Z_{b-1}$, which are the cosets in $T$ of
$\mathbb{Z}^\nu$ with respect to $Z_0$. Then the reconstruction is applied in
sequence to each sublattice, and use is made of the fact that, if
$\ind\in Z_l$, there exists a bound to
$\gamma(\inda)$, $\lambda(\inda,\indb)$ and $\lambda(\indb,\inda)$ for $\inda\in
Z_l\backslash \{\ind\}$ and $\indb\in T\backslash \{\ind\}$ which does not change
after the application of $U_\ind$. In particular, this implies that
the bounds do not change for the already reconstructed sites in
$Z_l$. Neither does the reconstruction at site $\ind$ change, on
account of Lemma~4 of paper \cite{do2}, the value of
$\lambda(\ind,\inda)$ and $\lambda(\inda,\ind)$, for
$|\inda-\ind|>\range$. In this sense the reconstruction is local.

So, the
values of $\hat \gamma_{V}$ and $\hat\lambda_{V}$, after one
application of $U_\ind$, depend at most on the values of
$\gamma(\ind)$ and $\lambda(\inda,\ind)$ in $V\cup\partial_\range
V$. It is thus appearent that we can control the values of the
reconstructed quantities only in a set $V$ smaller than the set $V'$
on which we control $\gamma$ and $\lambda$ initially. In particular,
$V$ can be chosen so that $V'=V\cup \partial_\range V$. Therefore, for
any $V\subset S$, we define for $l=0,\ldots,b-1$ a nested sequence of
sets $V_{l+1}\equal V_l\cup \partial_\range V_l$, with $V_0\equal V$
and the operator $U_V$, as $U_V=U_{V_0\cap Z_0}\circ \ldots\circ
U_{V_{b-1}\cap Z_{b-1}}$, and notice that (see the above remarks) the order in 
which the sites in $V_l\cap Z_l$ are chosen does not matter. Then,
after the application of $U_V$, one has that
$$
\left(\begin{array}{c}
\hat\gamma_V\\ 
\hat \lambda_V
\end{array}\right) \le 
A
\left(\begin{array}{c}
\gamma_{V_b}\\ 
\lambda_{V_b}
\end{array}\right)\ ,
$$
for the same matrix $A$ appearing in (\ref{eq:matrice_completo}). This
concludes the proof.

\end{document}